\PassOptionsToPackage{numbers,sort&compress}{natbib}
\documentclass[final,5p,times,twocolumn]{elsarticle}

\usepackage{amssymb}
\usepackage{amsmath}
\usepackage{graphicx}
\usepackage{xcolor}
\usepackage{bm}        

\usepackage{slashed}
\setcitestyle{numbers,sort&compress}
\usepackage{hyperref}
\usepackage{ctable}

\journal{Physics Letter B}

\begin{document}

\begin{frontmatter}



\title{Search for a generic heavy Higgs at the LHC}


\author[addr1,addr3]{Xin Chen}
\ead{xin.chen@cern.ch}

\author[addr1]{Yue Xu}
\ead{yue.xu@cern.ch}

\author[addr5]{Yongcheng Wu}
\ead{ycwu@physics.carleton.ca}

\author[addr1,addr3]{Yu-Ping Kuang}
\author[addr1,addr3]{Qing Wang}
\author[addr1]{Hang Chen}
\author[addr6]{Shih-Chieh Hsu}
\author[addr1,addr3]{Zhen Hu}
\author[addr6]{Congqiao Li}

\address[addr1]{Department of Physics, Tsinghua University, Beijing 100084, China}
\address[addr3]{Center for High Energy Physics, Tsinghua University, Beijing 100084, China}
\address[addr5]{Ottawa-Carleton Institute for Physics, Carleton University, Ottawa, Ontario K1S 5B6, Canada}
\address[addr6]{Department of Physics, University of Washington, Seattle, WA 98195, USA}

\begin{abstract}
A generic heavy Higgs has both dim-4 and effective dim-6 interactions with the Standard Model (SM) particles. The former has been the focus of LHC searches in all major Higgs production modes, just as the SM one, but with negative results so far. If the heavy Higgs is connected with Beyond Standard Model (BSM) physics at a few TeV scale, its dim-6 operators will play a very important role - they significantly enhance the Higgs momentum, and reduce the SM background in a special phase space corner to a level such that a heavy Higgs emerges, which is not possible with dim-4 operators only. We focus on the associated VH production, where the effect of dim-6 operators is the largest and the SM background is the lowest. Main search regions for this type of signal are identified, and substructure variables of boosted jets are employed to enhance the signal from backgrounds. The parameter space of these operators are scanned over, and expected exclusion regions with 300 fb$^{-1}$ and 3 ab$^{-1}$ LHC data are shown, if no BSM is present. The strategy given in this paper will shed light on a heavy Higgs which may be otherwise hiding in the present and future LHC data.

\end{abstract}


\end{frontmatter}

\section{Effective couplings of a heavy Higgs}

It is not very natural that the SM has only one fundamental scalar field - the Higgs field. If Nature really chooses this way, there must be something else unknown to us as yet. An alternative, and natural, way is that the 125 GeV Higgs boson discovered at the LHC \cite{ATLAS-Higgs}-\cite{CMS-Higgs} may be the lightest Higgs scalar field, among many that have yet to be found. Heavy Higgs particles are predicted in many BSM theories, such as the two-Higgs-doublet models, the minimal supersymmetric extension of the SM, and the left-right symmetric models. In a multiple Higgs field theory, the original Higgs fields are $\Phi_1,\Phi_2,\cdots\cdots$~\footnote{In general, they can be in any allowed SU(2)$_L$ representations. For simplicity, we will just illustrate the case where all fields are doublet.}. The multi-Higgs potential will cause mixing among them to form the mass eigenstates. Let $\Phi_h$ and $\Phi_H$ be the two doublets containing the lightest ($h$) and next to lightest ($H$) neutral Higgs respectively. 
The couplings to the SM gauge bosons will be scaled due to the mixing compared with SM gauge coupling.
At leading order, the dim-4 operators can be written as
%
\begin{align}
\label{equ:dim4}
\mathcal{L}^{(4)}_{hWW} & = 
\rho_h g m_W h W^\mu W_\mu, \nonumber \\
\mathcal{L}^{(4)}_{hZZ} & = 
\rho_h \frac{g m_W}{2\cos^2\theta_W} h Z^\mu Z_\mu, \nonumber \\
\mathcal{L}^{(4)}_{HWW} & = 
\rho_H g m_W H W^\mu W_\mu, \nonumber \\
\mathcal{L}^{(4)}_{HZZ} & =  
\rho_H \frac{g m_W}{2\cos^2\theta_W} H Z^\mu Z_\mu,
\end{align}
where $\theta_W$ is the weak mixing angle, $m_W$ the $W$ boson mass,  $\rho_h$ and $\rho_H$ are the scaling factors. In the simplest 2HDM example, we will have $\rho_h=\cos(\beta-\alpha)$, $\rho_H=\sin(\beta-\alpha)$.


For a SM-like light Higgs, $\rho_h$ is not far away from 1. Generally, for a heavy Higgs $H$, there could also be dim-6 effective operators which is related to an even higher energy scale BSM physics~\cite{Kuang}:
\begin{equation}
\label{equ:dim6_general}
\mathcal{L}^{(6)}_{HVV} = \sum_{n} \frac{f_n}{\Lambda^2} \mathcal{O}_n ,
\end{equation}
where $\Lambda$ is the scale below which the effective Lagrangian holds. It is set to 5 TeV in this work, since BSM at this scale is hard to be probed directly in general. Similar operators also exist for the SM Higgs $h$. As mentioned in \cite{Kuang}, the dim-6 operators that are not constrained by precision electroweak (EW) data and relevant for the heavy Higgs are
\begin{eqnarray}
\label{equ:op_dim6}
\mathcal{O}_{WW} & = & \Phi^{\dagger}_H \hat{W}_{\mu\nu} \hat{W}^{\mu\nu}  \Phi_H, \nonumber \\
\mathcal{O}_{BB} & = & \Phi^{\dagger}_H \hat{B}_{\mu\nu} \hat{B}^{\mu\nu}  \Phi_H, \nonumber \\
\mathcal{O}_{W} & = & \left(D_\mu\Phi_H\right)^\dagger \hat{W}^{\mu\nu} \left(D_\nu\Phi_H\right), \nonumber \\
\mathcal{O}_{B} & = & \left(D_\mu\Phi_H\right)^\dagger \hat{B}^{\mu\nu} \left(D_\nu\Phi_H\right), 
\end{eqnarray}
where $\hat{B}_{\mu\nu}=i\frac{g'}{2}B_{\mu\nu}$ and $\hat{W}_{\mu\nu}=i\frac{g}{2}\sigma^a W^a_{\mu\nu}$. After the EW symmetry breaking, the effective Lagrangian terms involving the heavy Higgs and $W$/$Z$ bosons are
%
\begin{align}
  \label{equ:dim6}
  \mathcal{L}^{(6)}_{HWW} & = g m_W \frac{f_W}{2\Lambda^2}\left( W_{\mu\nu}^+ W^{-\mu} \partial^\nu H + h.c. \right) \nonumber \\
  & \quad -g m_W \frac{f_{WW}}{\Lambda^2} W_{\mu\nu}^+ W^{-\mu\nu} H , \nonumber \\
  \mathcal{L}^{(6)}_{HZZ} & = g m_W \frac{c^2 f_W + s^2 f_B}{2c^2 \Lambda^2}  Z_{\mu\nu} Z^\mu \partial^\nu H \nonumber \\
  & \quad - g m_W \frac{c^4 f_{WW} + s^4 f_{BB}}{2c^2 \Lambda^2}  Z_{\mu\nu} Z^{\mu\nu} H ,
\end{align}

where $s=\sin\theta_W$ and $c=\cos\theta_W$. Similar terms exist for $H\gamma\gamma$ and $HZ\gamma$ vertices, but relatively suppressed by $s$ and $s^2$\footnote{We will postpone the $H\gamma\gamma$ and $HZ\gamma$ analyses to a later work.}. In addition, to simplify the parameter space, we also neglect terms of $O(s^2)$ and $O(s^4)$ in Eq. \ref{equ:dim6}, which involve coefficients $f_B$ and $f_{BB}$, as done in Ref. \cite{Kuang}. 

\section{Search for a generic heavy Higgs at the LHC}

\subsection{Main search channels}

Heavy Higgs have been intensively searched for at the LHC in the $H\to ZZ\to 4\ell$ decay \cite{ATLAS-4lep}-\cite{CMS-4lep} and in the diboson final state \cite{ATLAS-diboson}-\cite{CMS-diboson}, with negative results so far. The main production channel is gluon-gluon fusion (ggF). It is reasonable to assume that the Yukawa coupling between the heavy Higgs and fermions is small, or the Higgs is even fermi-phobic, so that it can escape the direct detection in the ggF channel. The remaining production channels are associated VH (V=$W/Z$) and Vector Boson Fusion (VBF), which only involve the interactions between heavy Higgs and $W/Z$ bosons. 

Different from \cite{Kuang,Kuang2} where final states with just one lepton and multiple jets are used, we start with at least two leptons. Specifically, the following channels are investigated in this work:
\begin{itemize}
\item $VH\to \ell^+ \ell^- j j j j$, where the two leptons ($\ell$) are of opposite sign (OS) charge and same flavor ($e$ or $\mu$), which originate from a $Z$ boson decay, and a number of jets denoted by $j$. This is called the 2$\ell$ OS channel.
\item $VH\to \ell \nu \ell^+ \ell^- j j$, where one pair of lepton originate from a $Z$ boson decay. This is called the 3$\ell$ channel.
\item $VH\to \ell^\pm \nu \ell^\pm \nu j j$, where the two leptons are of same sign (SS) charge. This final state originates from the $WH\to W^\pm W^\pm W^\mp$ decay mode. This is called the 2$\ell$ SS channel.
\end{itemize}


In the 2$\ell$ SS and 3$\ell$ channel, the heavy Higgs mass can be reconstructed. It is not possible in the 2$\ell$ SS channel, but the signal sensitivity is the highest in this channel due to the low background. In principle, the channel $WH\to W^\pm W^\pm W^\mp \to 3\ell 3\nu$ can be also used, and we can additionally require no pair of leptons with OS charge and same flavors to suppress the $Z$+$X$ background. However, the signal yield of this channel is only about ~10\% of that in the  2$\ell$ SS channel. Therefore, we do not consider this channel here. We also checked the $VH\to\ell\nu j j j j$ channel as used in \cite{Kuang,Kuang2}. Although the signal yield is about a factor 10 larger than in the 2$\ell$ OS, the $W$+jets background is also about ten times larger than $Z$+jets, and the $t\bar{t}\to W(\ell\nu)b$+jets can be another major background even after the $b$-jet veto\footnote{On the other hand, the $t\bar{t}$ background is severely suppressed by the $Z$ mass window cut in the 2$\ell$ OS channel.}. Therefore, the sensitivity of  $VH\to\ell\nu j j j j$ is not expected to be much higher than 2$\ell$ OS, which is the least sensitive in the three channels considered  in this work.

\begin{figure}[!htb]
\centering
\includegraphics[width=0.45\textwidth]{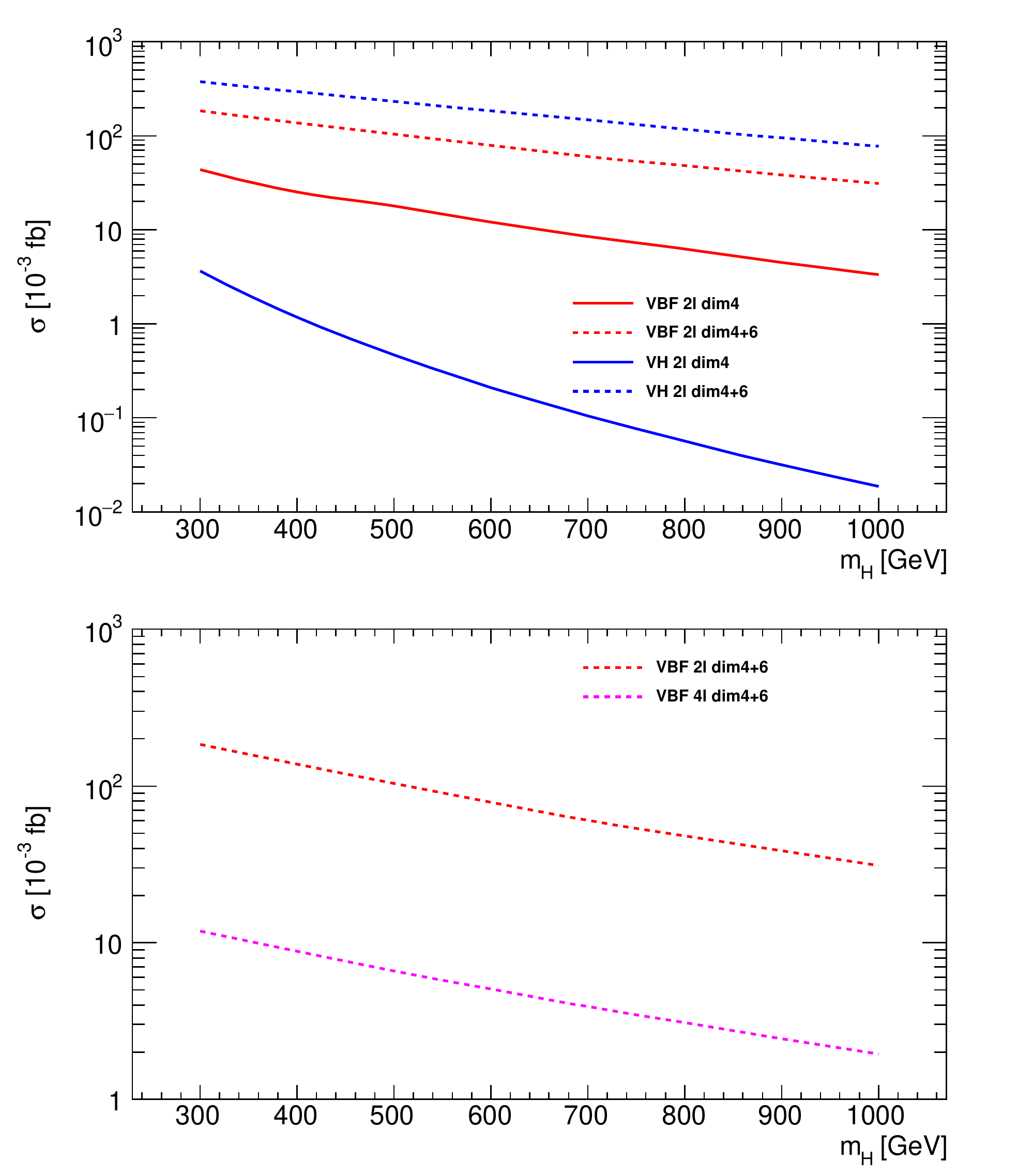}
\put(-180, 168){\textbf{(a)}}
\put(-180, 30){\textbf{(b)}}
\caption{ The LO cross sections of signal as a function of the heavy Higgs mass in the $2\ell$ and $4\ell$ channels. Two types of signals are compared. Dim4: $\rho_H=0.05$, $f_W=f_{WW}=0$; Dim4+6: $\rho_H=0.05$, $f_W=f_{WW}=50$. In (a), the VBF (VH) process is $pp\to Hjj, H\to\ell\ell jj$ ($pp\to VH\to \ell\ell+4j$). In (b), the VBF $4\ell$ process is $pp\to Hjj, H\to 4\ell$. }
\label{fig:vbf_vh_xs}
\end{figure}

\begin{figure}[!htb]
\centering
\includegraphics[width=0.45\textwidth]{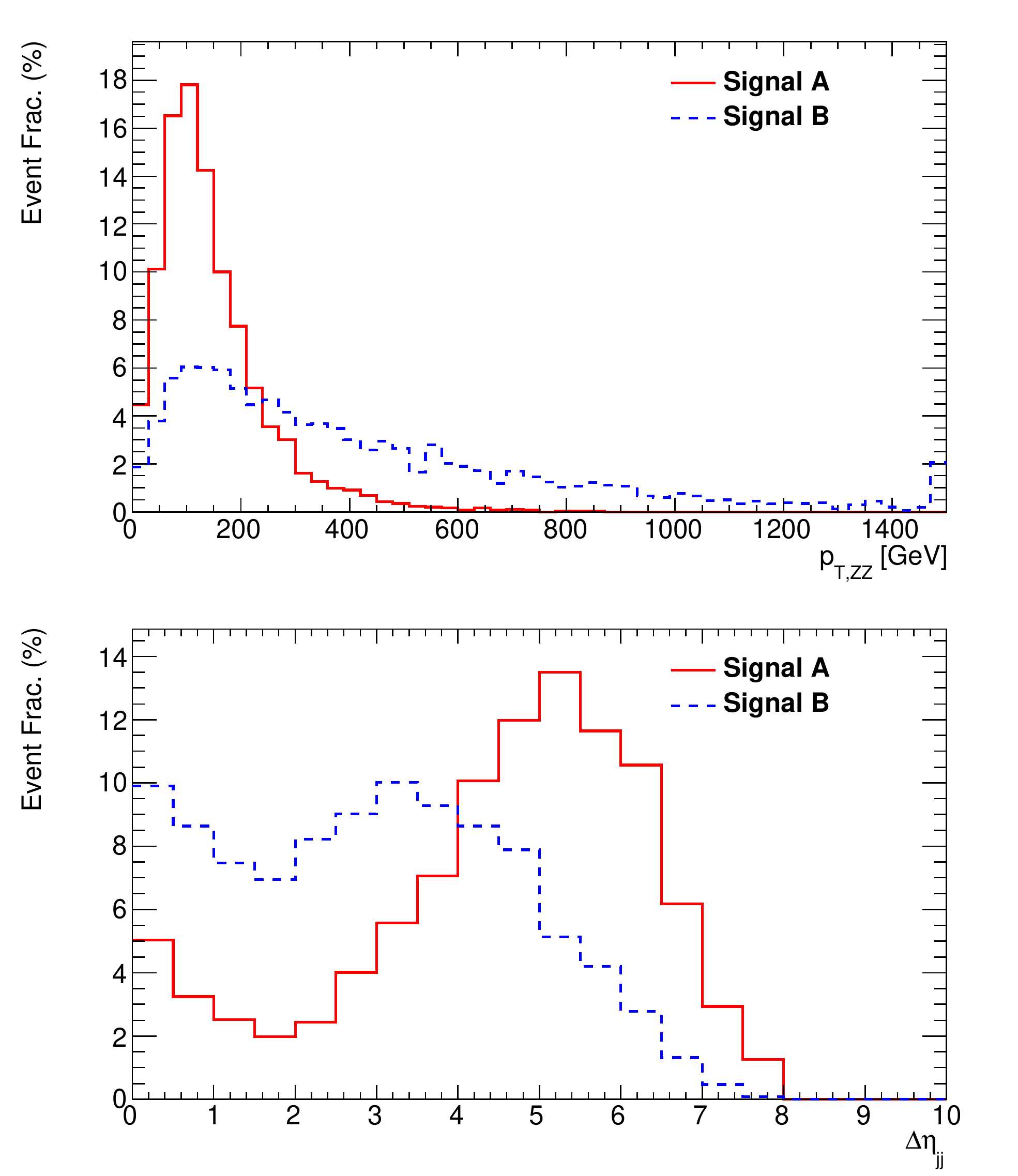}
\caption{The distributions of heavy Higgs $p_T$ (top) and $\Delta\eta_{jj}$ between the two leading jets (bottom) for two benchmark VBF signals with the $H \to ZZ \to 4\ell$ decay. A: $m_H=600$ GeV, $\rho_H=0.1$, $f_W=f_{WW}=0$, with $\sigma=3.1\times 10^{-3}$ fb; B: $m_H=600$ GeV, $\rho_H=0.05$, $f_W=f_{WW}=50$, with $\sigma=5.1\times 10^{-3}$ fb.}
\label{fig:vbf_sig}
\end{figure}

In general, the cross section of VBF is about an order of magnitude higher than VH in the high mass region, so it seems that VBF is the best channel to look for a heavy Higgs, and to suppress backgrounds by the presence of leptons in the final state, the decay modes of $H \to ZZ \to \ell\ell jj$ and $H \to ZZ \to 4\ell$ can be used. However, the former is accompanied by large SM backgrounds, and the yield of the latter is too small to be detected in the high momentum region. Therefore, we focus on the VH production mode, with the heavy Higgs decaying into two $W/Z$ bosons, and final states with at least two leptons from the three bosons' decays, as listed above. Figure \ref{fig:vbf_vh_xs} shows the leading order (LO) cross section of signal with different parameters as a function of the heavy Higgs mass. It is evident that when dim-6 operators are present, both VH and VBF production cross sections increase significantly, and VH increases much more than the VBF process.
In addition, some traditional VBF variables such as $\Delta\eta_{jj}$ may stop working for dim-6 operators. A comparison of two benchmark signals in the VBF $H \to ZZ \to 4\ell$ channel is made in Fig. \ref{fig:vbf_sig}, one with and another without the dim-6 operators. The presence of these operators enhances the Higgs $p_T$, but also makes $\Delta\eta_{jj}$ background-like. Both signals have a yield of no more than 0.5 event at 300 $\text{fb}^{-1}$ after object selection cuts, since their cross sections are already at $\mathcal{O}(10^{-3}-10^{-2})$ fb level before any  detector level cuts, as indicated in the caption of Fig. \ref{fig:vbf_vh_xs}(b). As a result, the $4\ell$ channel significances are much lower than $2\ell$ and $3\ell$, and we do not included it in the final result.
%

The extra derivatives in Eq. \ref{equ:dim6} will not only increase the heavy Higgs process cross section substantially, but also make the heavy Higgs and associated boson have high momenta. Combined with the large Higgs mass, this means that all three bosons present in the process are boosted, which leads to boosted boson jets in the final state. We can use both the high $p_T$ and substructure features of these jets to suppress the backgrounds. For large $\rho_H$, the contribution from off-shell $VH^*$ production can be also sizable.

\begin{figure}[!htb]
\centering
\includegraphics[width=0.45\textwidth]{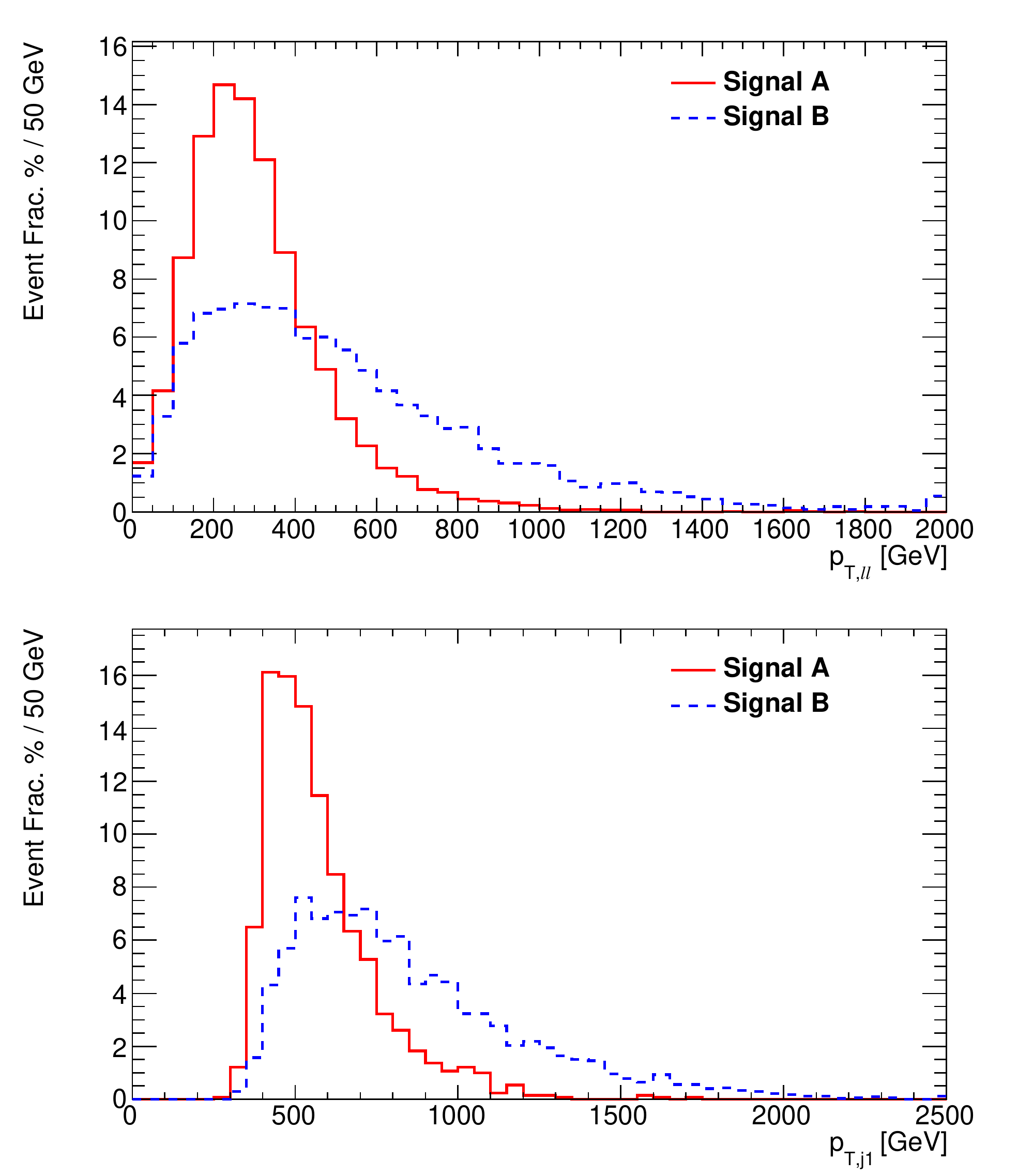}
\caption{The distributions of the dilepton $p_T$ (top) and leading jet $p_T$ whose jet mass is consistent with a $W/Z$ boson (bottom) for two benchmark VH signals in the $2\ell$ channel. A: $m_H=600$ GeV, $\rho_H=0.1$, $f_W=f_{WW}=0$, with $\sigma=8.5\times 10^{-4}$ fb; B: $m_H=600$ GeV, $\rho_H=0.05$, $f_W=f_{WW}=50$, with $\sigma=0.18$ fb.}
\label{fig:vh_sig}
\end{figure}

The dilepton and leading jet (with $70~\text{GeV}< m_j <150~\text{GeV}$) $p_T$ for VH production in the $2\ell$ channel are shown in Fig. \ref{fig:vh_sig}.
Indeed, the bosons in signals with dim-6 operators have higher $p_T$'s.
%

\subsection{Simulation of signal and background events}

The effective interactions in Eq. \ref{equ:dim6} are modeled by {\sc FeynFules} \cite{Alloul:2013bka} and passed to {\sc MadGraph5} \cite{MG5} for the heavy Higgs production and decay, and the partons are showered and hadronized by {\sc Pythia8} \cite{Pythia8}. The free parameters that can be set freely in this model are $m_H$, $\rho_H$, $f_W$ and $f_{WW}$. In the $2\ell$ channels, single boson background of $\ell\ell$ plus up to four QCD partons and diboson process $\ell\ell+jj$ (where the two $j$'s come from EW vertices) plus up to two QCD partons, are generated at the matrix element level with {\sc MadGraph5} and matched to the parton showers with the MLM method~\cite{MLM}. The triboson process $\ell\ell+4j$ (where the four $j$'s come from EW vertices), $t\bar{t}$ with $t\to \ell\nu b$, and $t\bar{t}V$ with two leptons from the decays of top or $V(W/Z)$, are also generated with {\sc MadGraph5}\footnote{It is worthwhile to note that ATLAS sees evidence of the SM triboson process with partial 13 TeV data~\cite{ATLAS-triboson}.}. In the $3\ell$ channel, the diboson process $3\ell+\nu$ plus up to two QCD partons are generated with {\sc MadGraph5} and matched to the parton showers with the MLM method. The triboson process $3\ell+\nu+jj$ (where the two $j$'s come from EW vertices), and $t\bar{t}V$ with three leptons in the final state are also generated with {\sc MadGraph5}. Our triboson events include the off-shell effect of the bosons, thus the SM VBF process is also included. All background samples are showered and hadronized by {\sc Pythia8} too. Both signals and backgrounds are generated at LO in QCD and EW at $\sqrt{s}=13$ TeV, and the PDF set NNPDF23LO \cite{Ball} is used for all the samples.

The events are afterwards passed through {\sc DELPHES} \cite{Delphes} simulating the detector response of the ATLAS detector \cite{ATLAS-det}. The tracking range is defined to be within $|\eta|<2.5$, where $\eta$ is the pseudorapidity. The electron tracking and identification efficiencies are $90-94\%$ ($71-77\%$) in the region $|\eta|\leq 1.5$ ($1.5<|\eta|<2.5$), and those for muons is $94\%$ ($83\%$) in $|\eta|\leq 1.5$ ($1.5<|\eta|<2.5$).
The jet and missing transverse energy ($E_T^\text{miss}$) are based on calorimeter measurements. The electromagnetic calorimeter resolutions are parametrized as $10.1\%\sqrt{E}\oplus 0.17\%E$ ($28.5\%\sqrt{E}\oplus 3.50\%E$) for $|\eta|\leq 3.2$ ($3.2<|\eta|<4.9$), while those for the hadronic calorimeter are $1.59~\text{GeV}\oplus 52.05\%\sqrt{E}\oplus 3.02\%E$, $70.6\%\sqrt{E}\oplus 5.00\%E$ and $100.0\%\sqrt{E}\oplus 9.42\%E$ for $|\eta|\leq 1.7$, $1.7<|\eta|\leq 3.2$ and $3.2<|\eta|<4.9$, respectively. The energy $E$ is all in GeV.

The minimum $p_T$ for an electron (muon) is 15 GeV (10 GeV). The normal jets are clustered with the anti-$k_{t}$ algorithm~\cite{antikt} with a cone parameter 0.4. To account for the boosted bosons and the Higgs, anti-$k_{t}$ fat jets with a cone parameter 1.0 are also used. If any jet (fat jet) overlaps with a lepton within $\Delta R<0.4$ ($\Delta R<1.0$)\footnote{The $\Delta R$ is defined as $\Delta R=\Delta\eta\oplus\Delta\phi$.}, this jet (fat jet) is removed in the event from consideration. A jet and a fat jet should also have $\Delta R>1.4$ to be considered as non-overlapping. The normal (fat) jets are required have $p_T>30$ GeV ($p_T>50$ GeV), and with $|\eta|<4.0$. 

\subsection{Search in the $2\ell$ OS and SS channels}

\begin{figure}[!htb]
\centering
\includegraphics[width=0.415\textwidth]{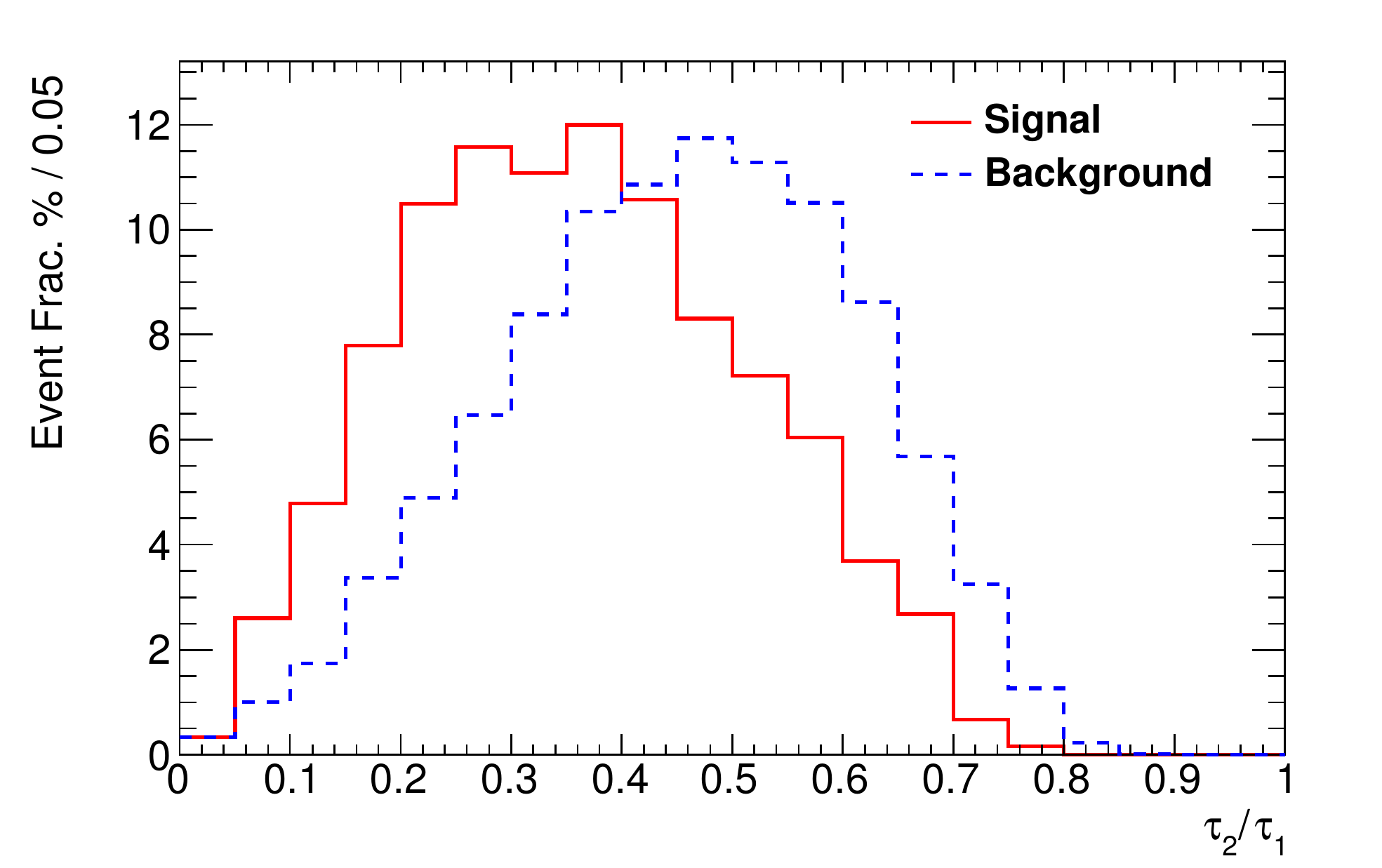}
\caption{The distribution of $\tau_{2}/\tau_{1}$ for the boson jets $j_{1,2}$ in region (2-4) of the $2\ell$ channel. The signal shown has the following parameters: $m_H=600$ GeV, $\rho_H=0.05$, $f_W=f_{WW}=50$.}
\label{fig:tau21_2lep}
\end{figure}

\begin{figure}[!htb]
\centering
\includegraphics[width=0.23\textwidth]{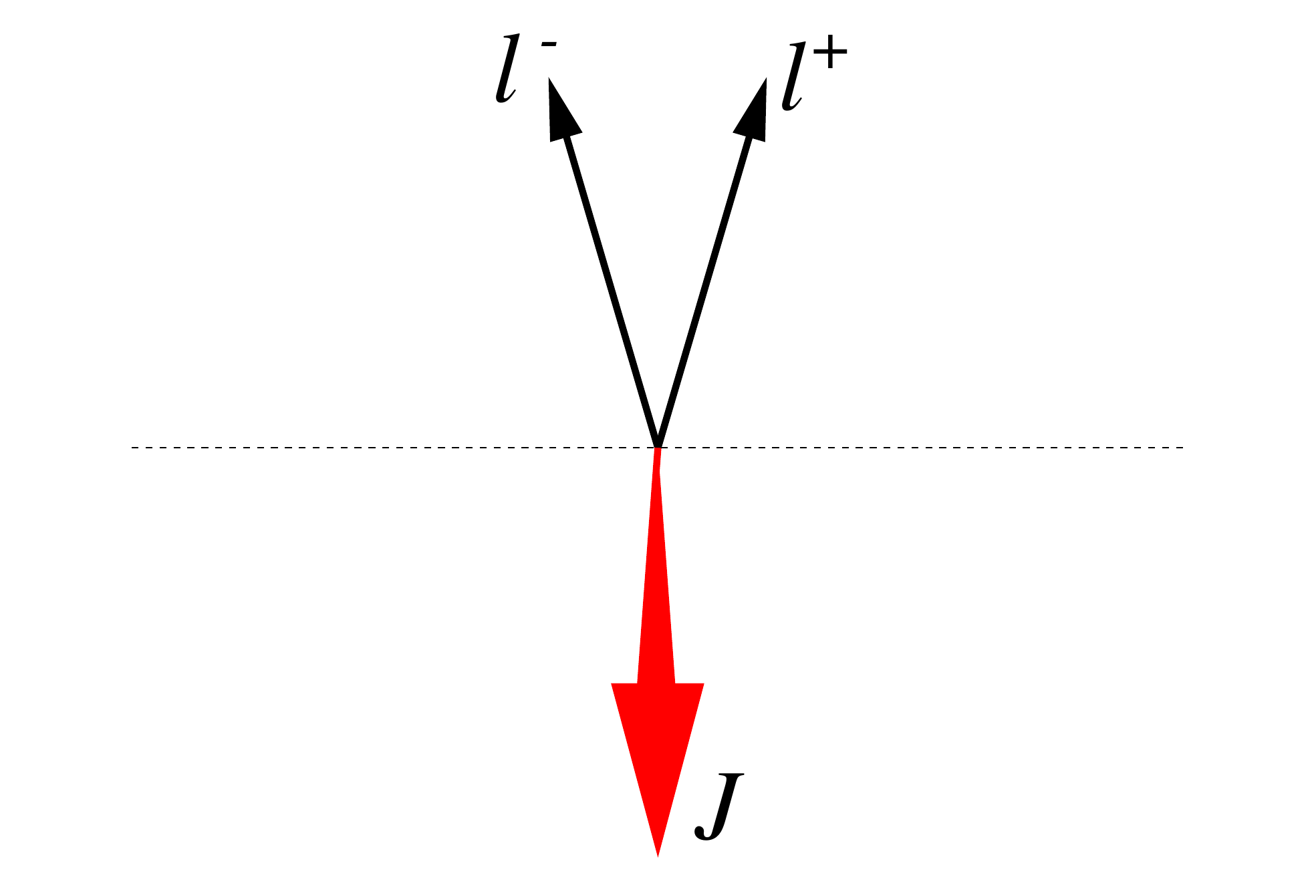}
\put(-21, 8){\textbf{(1)}}
\includegraphics[width=0.23\textwidth]{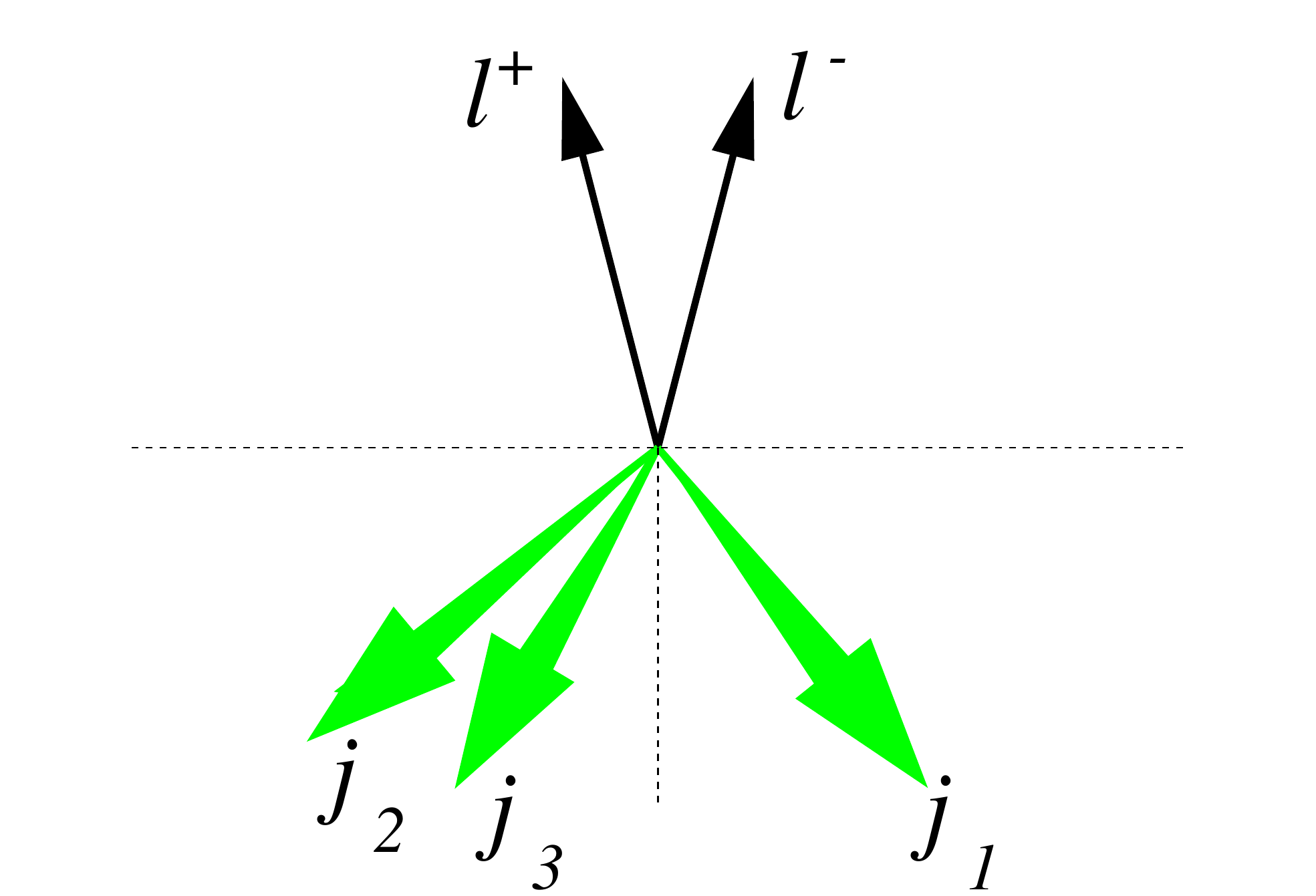}
\put(-21, 8){\textbf{(2)}}\\
\includegraphics[width=0.23\textwidth]{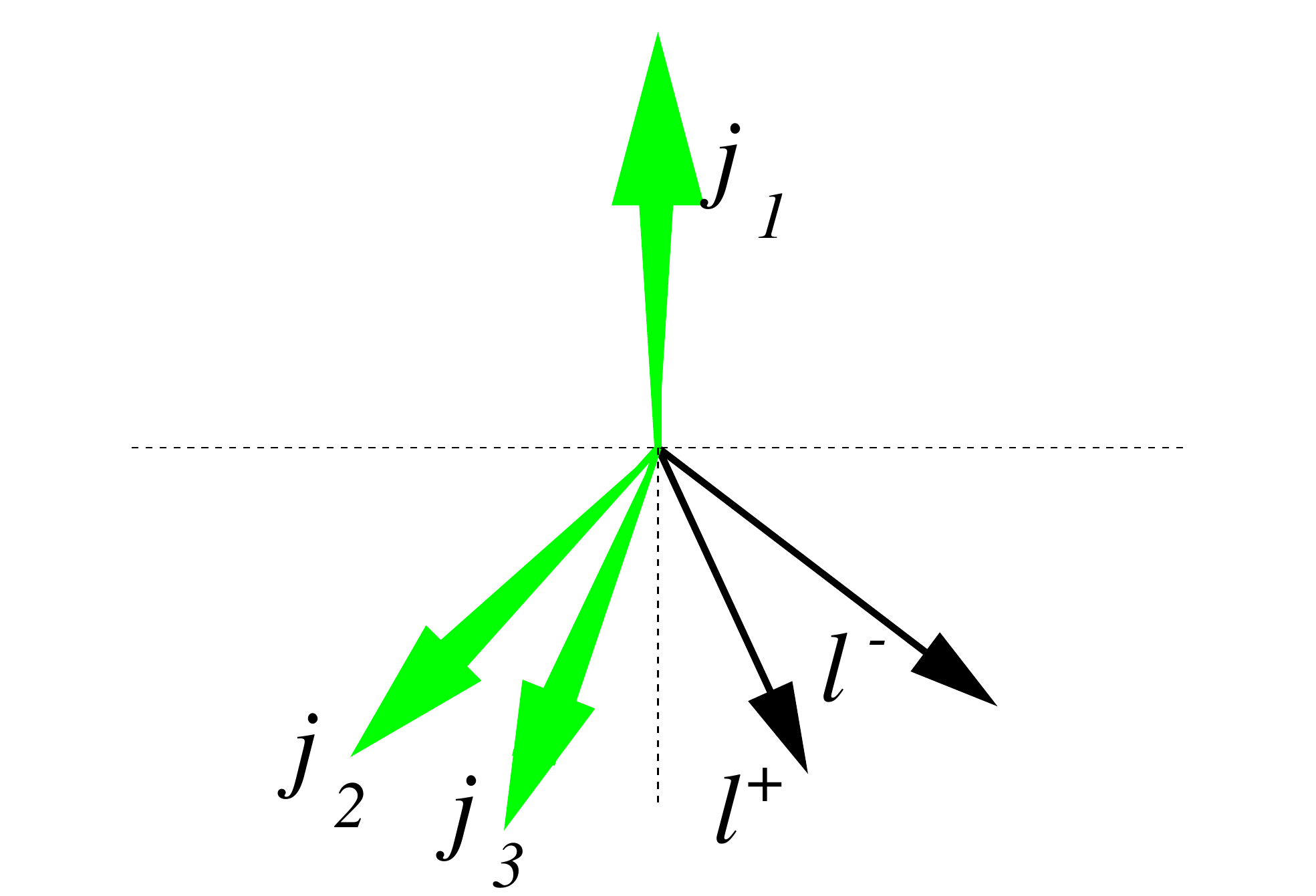}
\put(-21, 8){\textbf{(3)}}
\includegraphics[width=0.23\textwidth]{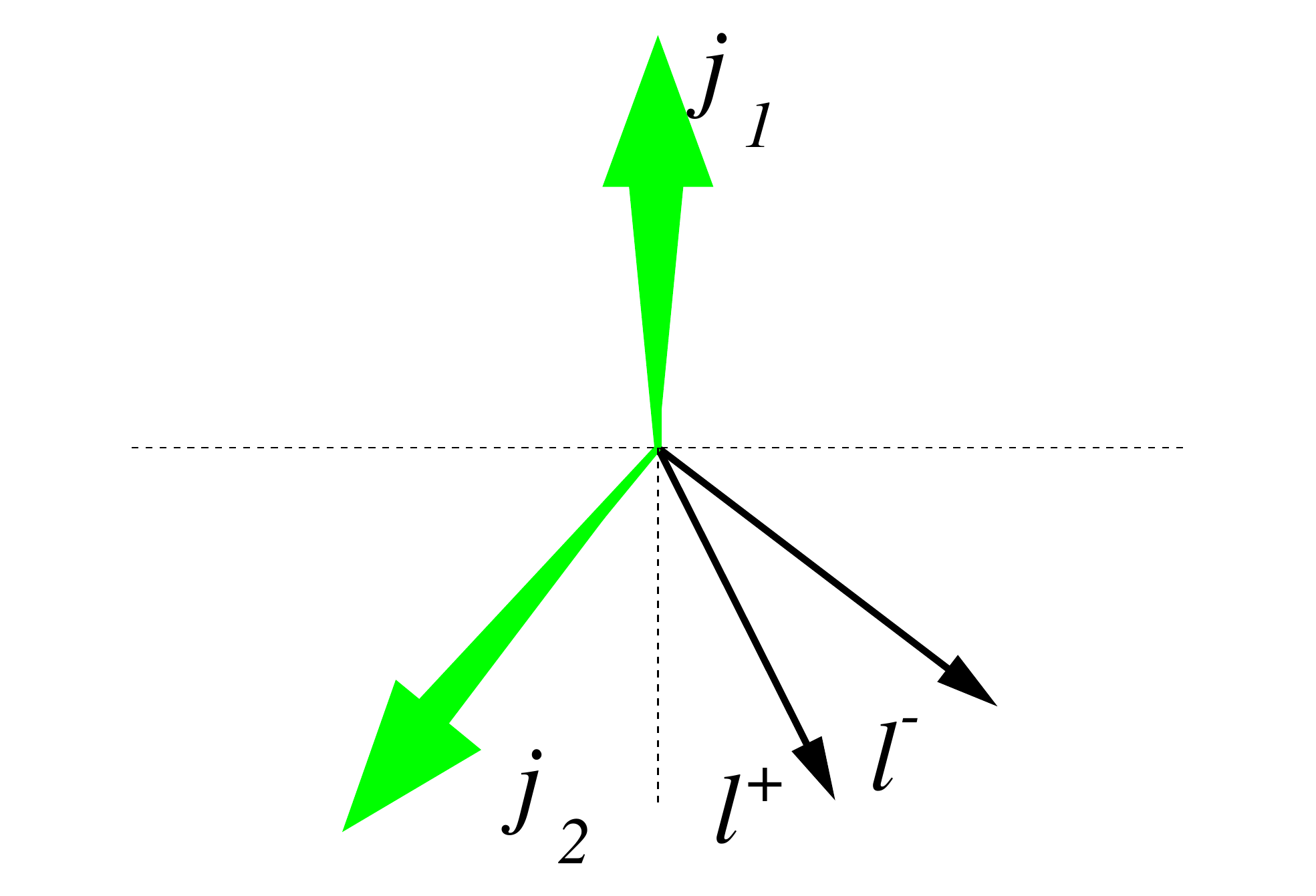}
\put(-21, 8){\textbf{(4)}}\\
\caption{The four signal regions defined in the $2\ell$ channel. The green (red) arrow denotes a normal (fat) jet.}
\label{fig:reg_2lep}
\end{figure}

To search for a heavy Higgs with boosted bosons in the $2\ell$ OS channel, four signal regions are defined as shown in Tab. \ref{tab:reg_2lep}. The event topology is characterized by a high momentum boson recoiling against two other bosons that come from a heavy Higgs decay, as schematically displayed in Fig. \ref{fig:reg_2lep}. In region (1), the associated $Z\to\ell\ell$ recoils against a high momentum Higgs decaying into four jets ($\ell\ell$ denotes the combined 4-vector of two leptons). The momentum is so high that the four jets form a fat jet (denoted by $J$). A a parameterless $k_t$ algorithm is run on the fat jet to exclusively cluster up to two subjets \cite{FastJet}. Exactly two such subjets are required, and each one's mass should be consistent with a vector boson. To further suppress backgrounds, the $N$-{\sc subjettiness} variables $\tau_{1,2}$ are used \cite{FastJet}. They are jet substructure variables calculated using exclusive $k_t$ axes, indicative of the subjet multiplicity in a parent jet. Similar topology to (1) exists in region (2), except that one boson from the heavy Higgs forms a boosted boson jet (a single normal jet denoted by $j_1$, which is leading in $p_T$), and the other with a lower $p_T$ splits into two normal jets ($j_2$ and $j_3$, $2^{nd}$ and $3^{rd}$ leading in $p_T$, and $j_{23}$ denotes the combined 4-vector of these two jets). In region (3) and (4), the leading jet is the associated boson. One boson from Higgs decay forms two jets (region 3) or a single jet (region 4), and the other decays into dilepton. The $\Delta R$ cuts are applied to imposed correct topologies in different regions. The distributions of $\tau_{2}/\tau_{1}$ for the boson jets $j_{1,2}$ in region (2-4) are shown in Fig. \ref{fig:tau21_2lep}.

For signals with $m_H=300$ GeV, the regions definitions are similar to $m_H=600$ GeV, but due to the lower Higgs mass, events have less number of boosted boson jets. A bit tighter mass window cut is applied on the bosons, and region (4) is removed due to poor signal significance. Conversely, in signal with $m_H=900$ GeV, events have much larger number of boosted boson jets, and the Higgs can hardly form a fat jet. As a result, region (1) and (3) are removed, and a new region similar to (2) but with $j_{23}$ replaced by a single normal jet is added. 

\begin{table}[!bt]
\centering
\caption{The signal region definitions for $m_H=600$ GeV in the $2\ell$ OS channel.}
\label{tab:reg_2lep}
\begin{tabular}{l | l}\specialrule{1.0pt}{1pt}{1pt}
 region (1) & region (2) \\ \hline
  \multicolumn{2}{c}{$80~\text{GeV}<m_{\ell\ell}<100~\text{GeV}$}\\ \hline
$p_T^{\ell\ell}>950~\text{GeV}$, & $p_T^{\ell\ell}>550~\text{GeV}$, \\
$p_T^J>750~\text{GeV}$, & $p_T^{j_1}>300~\text{GeV}$, \\ 
$N_{sj}=2$, & $70~\text{GeV}< m_{j_1} <150~\text{GeV}$, \\
$70~\text{GeV}< m_{sj_{1,2}} <150~\text{GeV}$, & $\tau_2^{j_1}/\tau_1^{j_1} < 0.40$, \\
$\tau_2^J/\tau_1^J < 0.45$ & $70~\text{GeV}< m_{j_{23}} <110~\text{GeV}$, \\ 
 & $p_T^{j_{23}}>150~\text{GeV}$, \\
 & $\Delta R(j_1,j_{23})<\Delta R(\ell\ell,j_1)$, \\
 & $\Delta R(j_1,j_{23})<\Delta R(\ell\ell,j_{23})$, \\
 & $p_T^{j_1+j_{23}}>550$ GeV \\ \specialrule{1.0pt}{1pt}{1pt}
  region (3) & region (4) \\ \hline
\multicolumn{2}{c}{$80~\text{GeV}<m_{\ell\ell}<100~\text{GeV}$, $p_T^{\ell\ell}>300~\text{GeV}$,} \\
\multicolumn{2}{c}{$p_T^{j_1}>700~\text{GeV}$, $70~\text{GeV}< m_{j_1} <150~\text{GeV}$} \\ \hline
$\tau_2^{j_1}/\tau_1^{j_1} < 0.60$, & $\tau_2^{j_1}/\tau_1^{j_1} < 0.52$, \\
$75~\text{GeV}< m_{j_{23}} <115~\text{GeV}$, & $p_T^{j_2}>250$ GeV, \\
$p_T^{j_{23}}>50~\text{GeV}$, & $70~\text{GeV}< m_{j_2} <150~\text{GeV}$, \\
$\Delta R(\ell\ell,j_{23})<\Delta R(j_1,\ell\ell)$, & $\tau_2^{j_2}/\tau_1^{j_2} < 0.52$, \\
$\Delta R(\ell\ell,j_{23})<\Delta R(j_1,j_{23})$, & $\Delta R(\ell\ell,j_2)<\Delta R(j_1,\ell\ell)$, \\
$p_T^{\ell\ell+j_{23}}>700$ GeV & $\Delta R(\ell\ell,j_2)<\Delta R(j_1,j_2)$, \\
 & $p_T^{\ell\ell+j_2}>700$ GeV\\
 \specialrule{1.0pt}{1pt}{1pt}
\end{tabular}
\end{table}

\begin{figure}[!htb]
\centering
\includegraphics[width=0.23\textwidth]{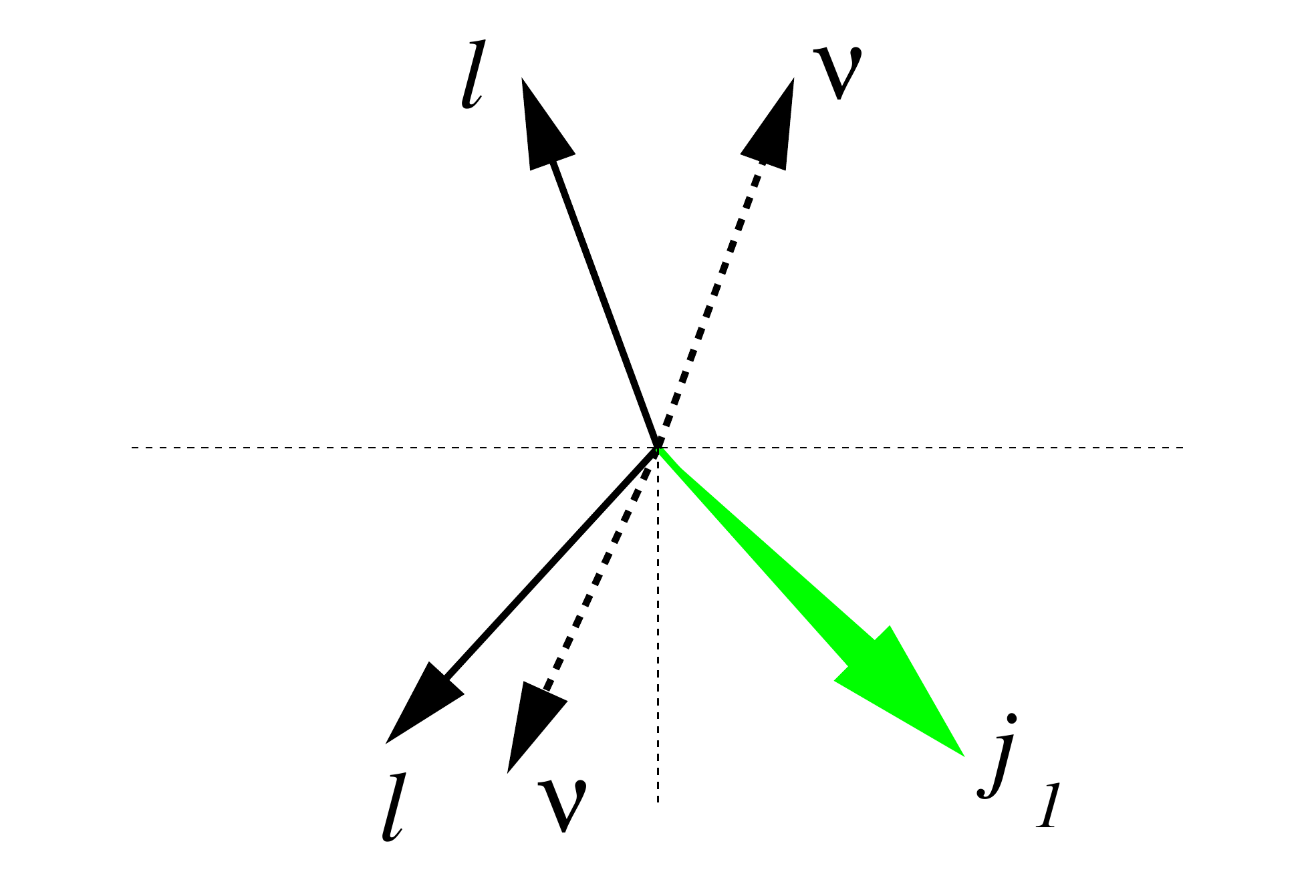}
\put(-21, 8){\textbf{(1)}}
\includegraphics[width=0.23\textwidth]{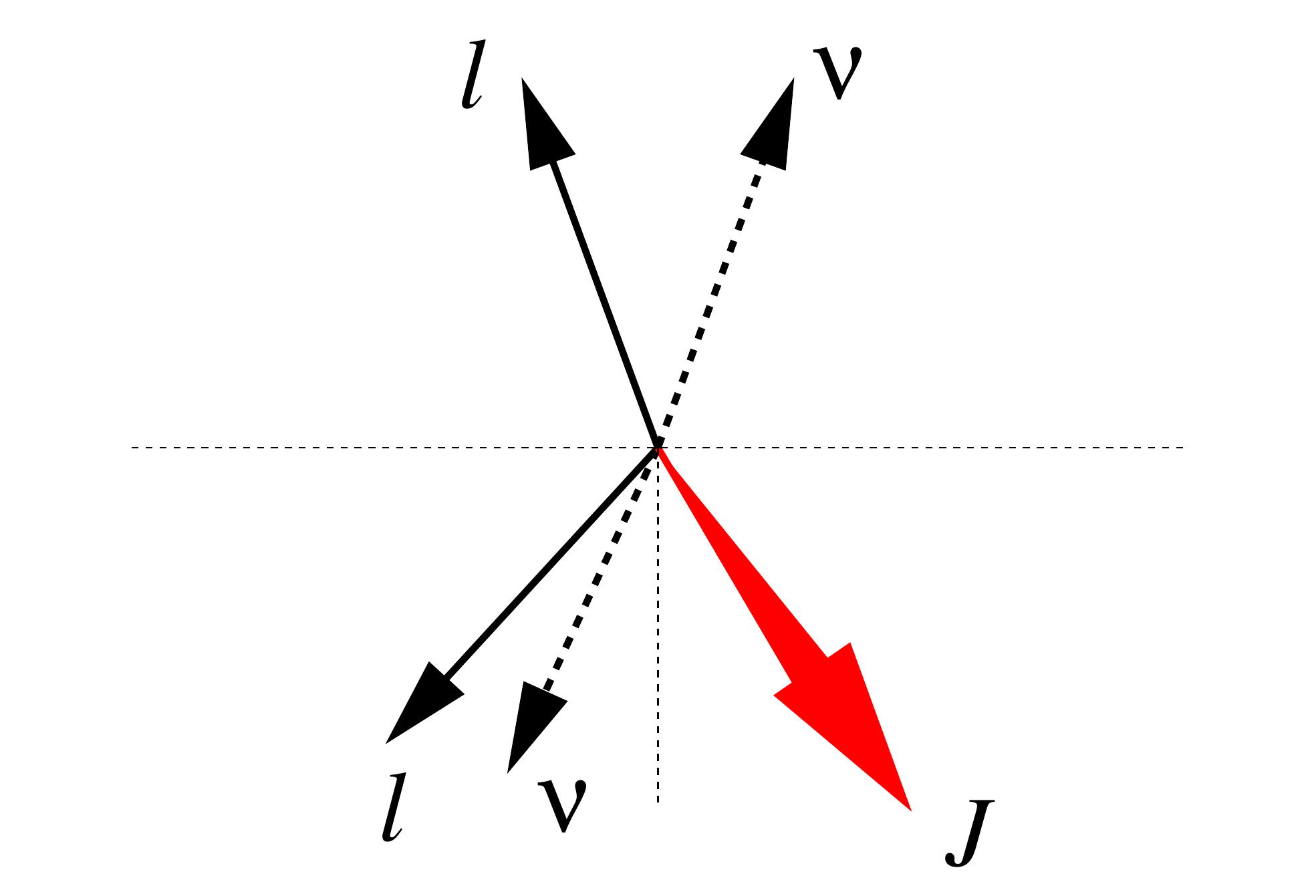}
\put(-21, 8){\textbf{(2)}}\\
\includegraphics[width=0.23\textwidth]{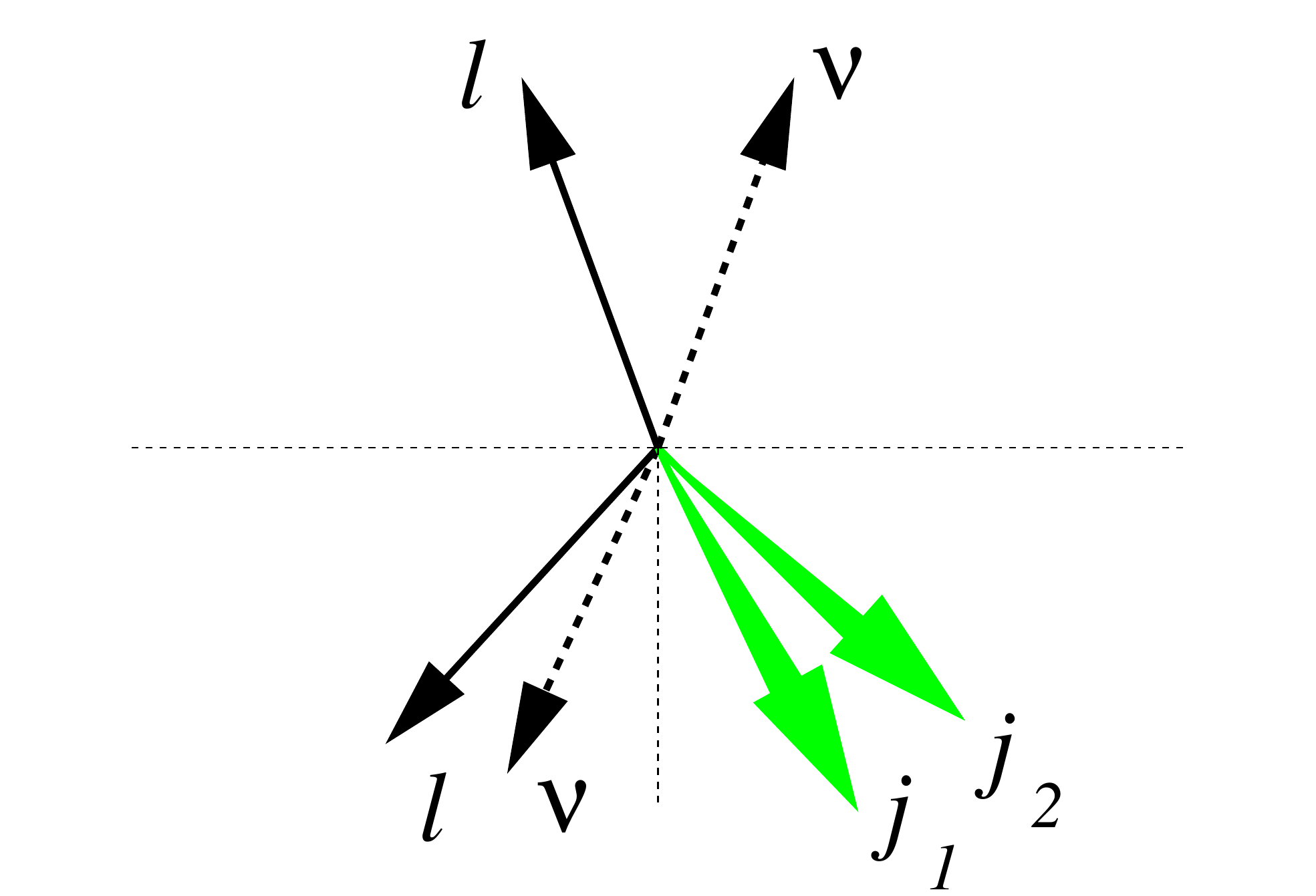}
\put(-21, 8){\textbf{(3)}}
\caption{The three signal regions defined in the $2\ell$ SS channel. The green (red) arrow denotes a normal (fat) jet.}
\label{fig:reg_2lep_SS}
\end{figure}

In the $2\ell$ SS channel, the $Z$+jets will be severely suppressed, and the main backgrounds consist of $WZ$+jets where one lepton from $Z$ missed the reconstruction or identification, $t\bar{t}V$ and triboson $WWW$. Three signal regions are similarly defined in Tab. \ref{tab:reg_2lep_SS}, and illustrated in Fig. \ref{fig:reg_2lep_SS}. Compared to the $2\ell$ OS channel, large $E_T^\text{miss}$ is required due to neutrinos, and $b$-jet veto is applied to suppress the $t\bar{t}V$ background. The minimum subleading lepton $p_T$ is 50 GeV, to suppress fake leptons which typically have lower $p_T$\footnote{The fake lepton background is not modeled in this work, and is expected to be small with the high lepton $p_T$ cut applied in this work. }. 

\begin{table}[!bt]
\centering
\caption{The signal region definitions for $m_H=600$ GeV in the $2\ell$ SS channel.}
\label{tab:reg_2lep_SS}
\begin{tabular}{l | l}\specialrule{1.0pt}{1pt}{1pt}
 region (1) & region (2) \\ \hline
  \multicolumn{2}{c}{$m_{\ell\ell}>300~\text{GeV}$, $p_T^{\ell\ell}>100~\text{GeV}$,}\\ 
  \multicolumn{2}{c}{$p_T^{\ell_1}>300~\text{GeV}$, $p_T^{\ell_2}>50~\text{GeV}$, }\\ 
  \multicolumn{2}{c}{$\Delta\phi_{\ell\ell}>2.0$, $E_T^\text{miss}>100~\text{GeV}$, }\\
  \multicolumn{2}{c}{no $b$-tagged jets }\\ \hline
$p_T^{j_1}>400~\text{GeV}$ & $p_T^{J}>100~\text{GeV}$, \\
& $\tau_2^J/\tau_1^J < 0.6$ \\ \specialrule{1.0pt}{1pt}{1pt}
\multicolumn{2}{l}{region (3)} \\ \hline
\multicolumn{2}{l}{$m_{\ell\ell}>400~\text{GeV}$, $p_T^{\ell\ell}>100~\text{GeV}$, }\\ 
\multicolumn{2}{l}{$p_T^{\ell_1}>450~\text{GeV}$, $p_T^{\ell_2}>50~\text{GeV}$, }\\ 
\multicolumn{2}{l}{$\Delta\phi_{\ell\ell}>1.6$, $E_T^\text{miss}>100~\text{GeV}$}\\ 
 \specialrule{1.0pt}{1pt}{1pt}
\end{tabular}
\end{table}

\subsection{Search in the $3\ell$ channel}

\begin{figure}[!htb]
\centering
\includegraphics[width=0.23\textwidth]{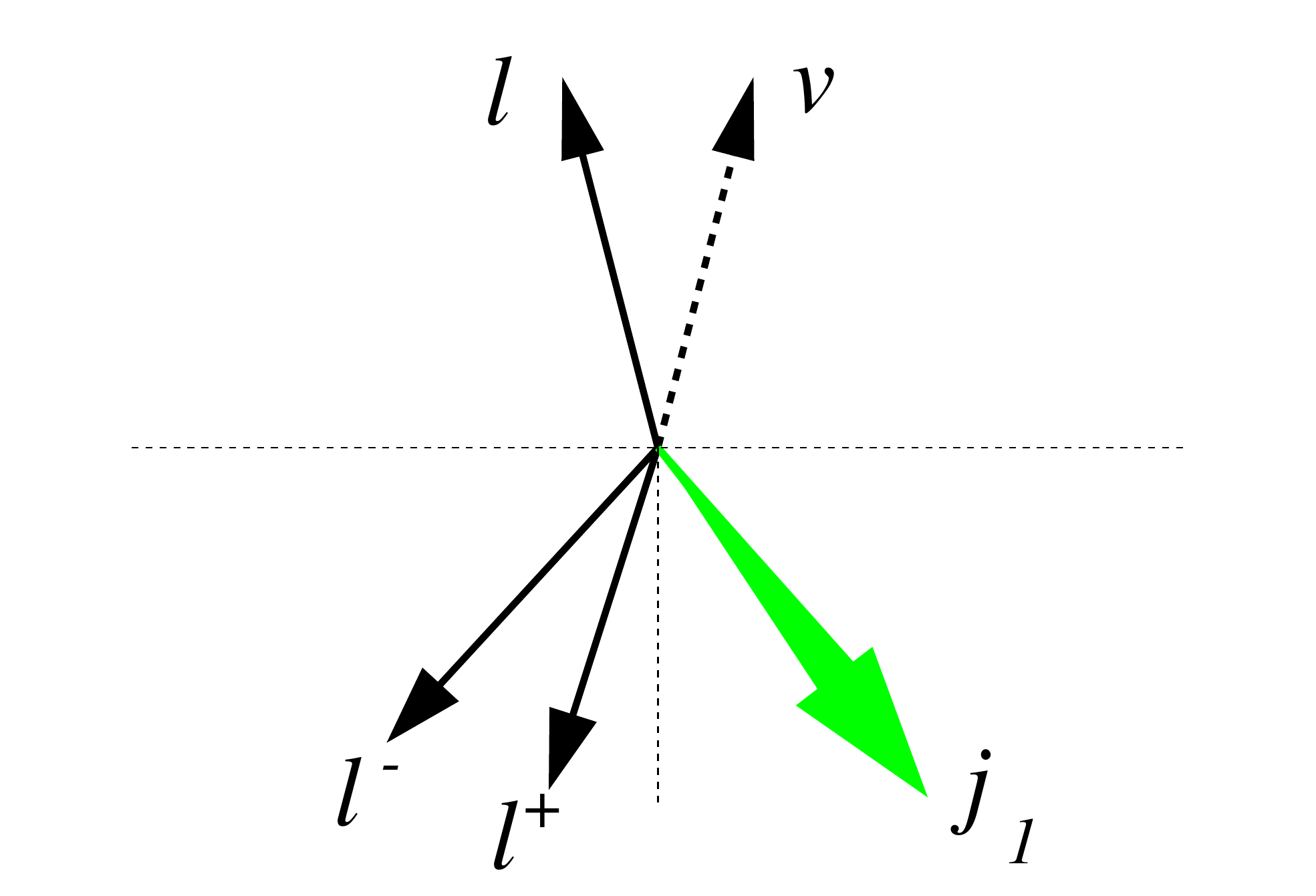}
\put(-21, 8){\textbf{(1)}}
\includegraphics[width=0.23\textwidth]{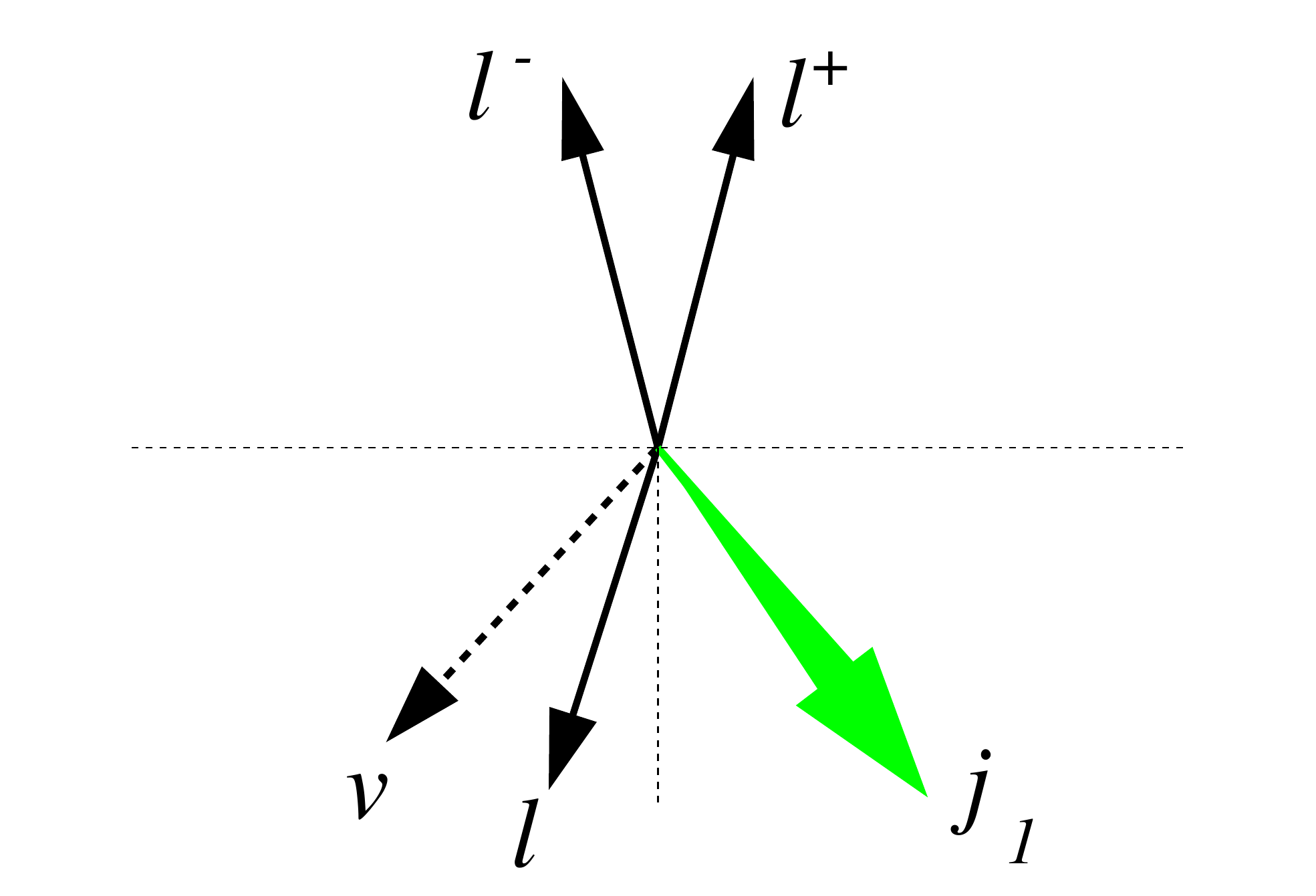}
\put(-21, 8){\textbf{(4)}}\\
\includegraphics[width=0.23\textwidth]{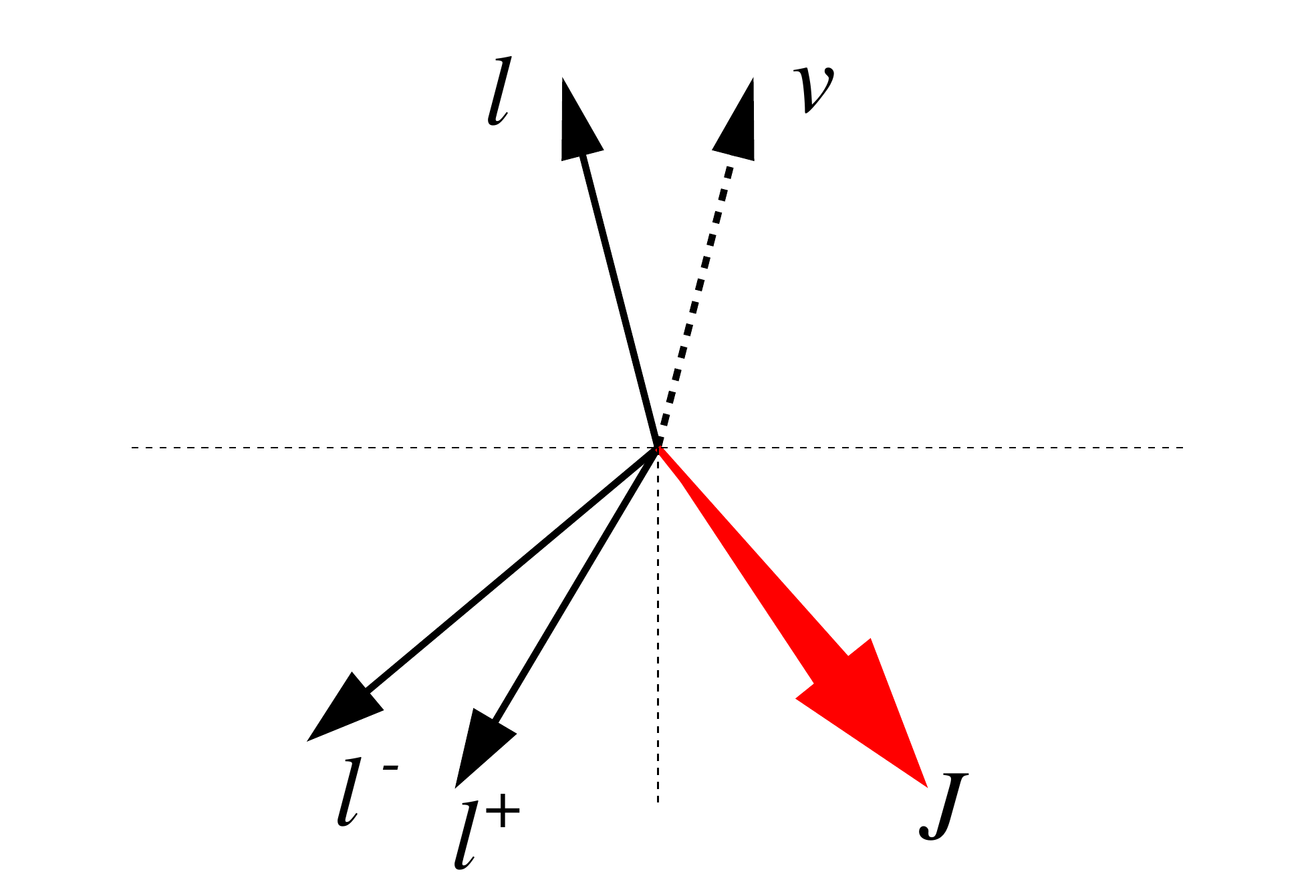}
\put(-21, 8){\textbf{(2)}}
\includegraphics[width=0.23\textwidth]{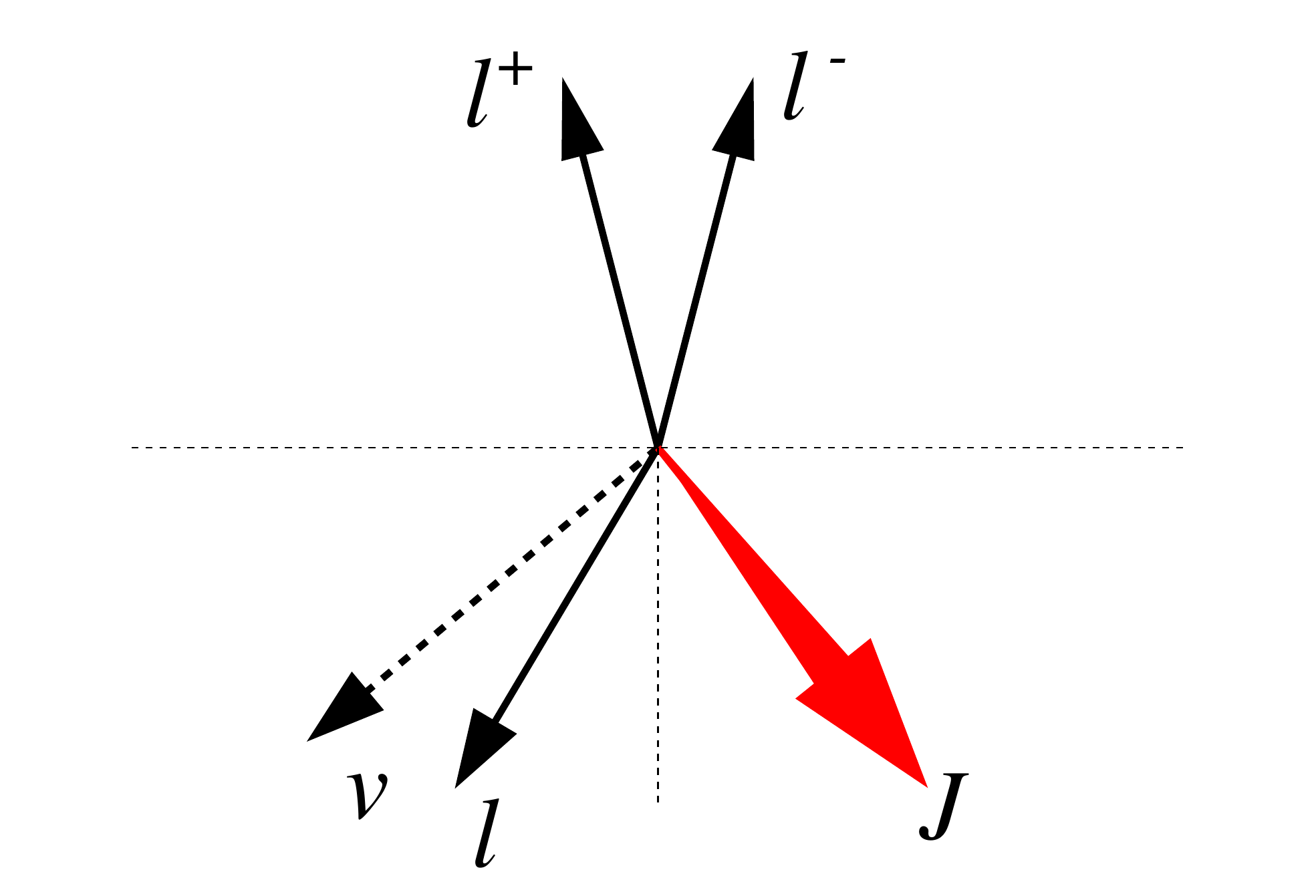}
\put(-21, 8){\textbf{(5)}}\\
\includegraphics[width=0.23\textwidth]{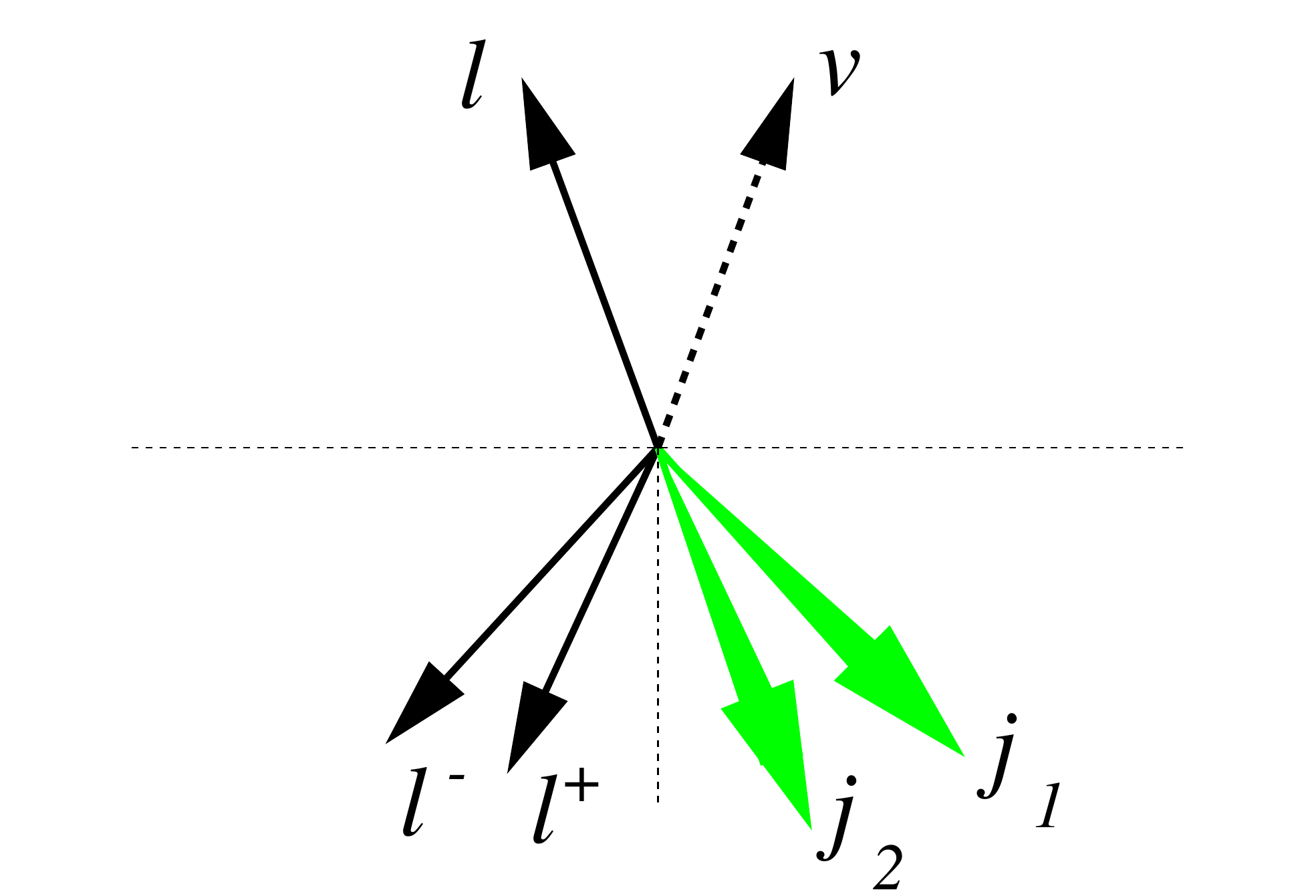}
\put(-21, 8){\textbf{(3)}}
\includegraphics[width=0.23\textwidth]{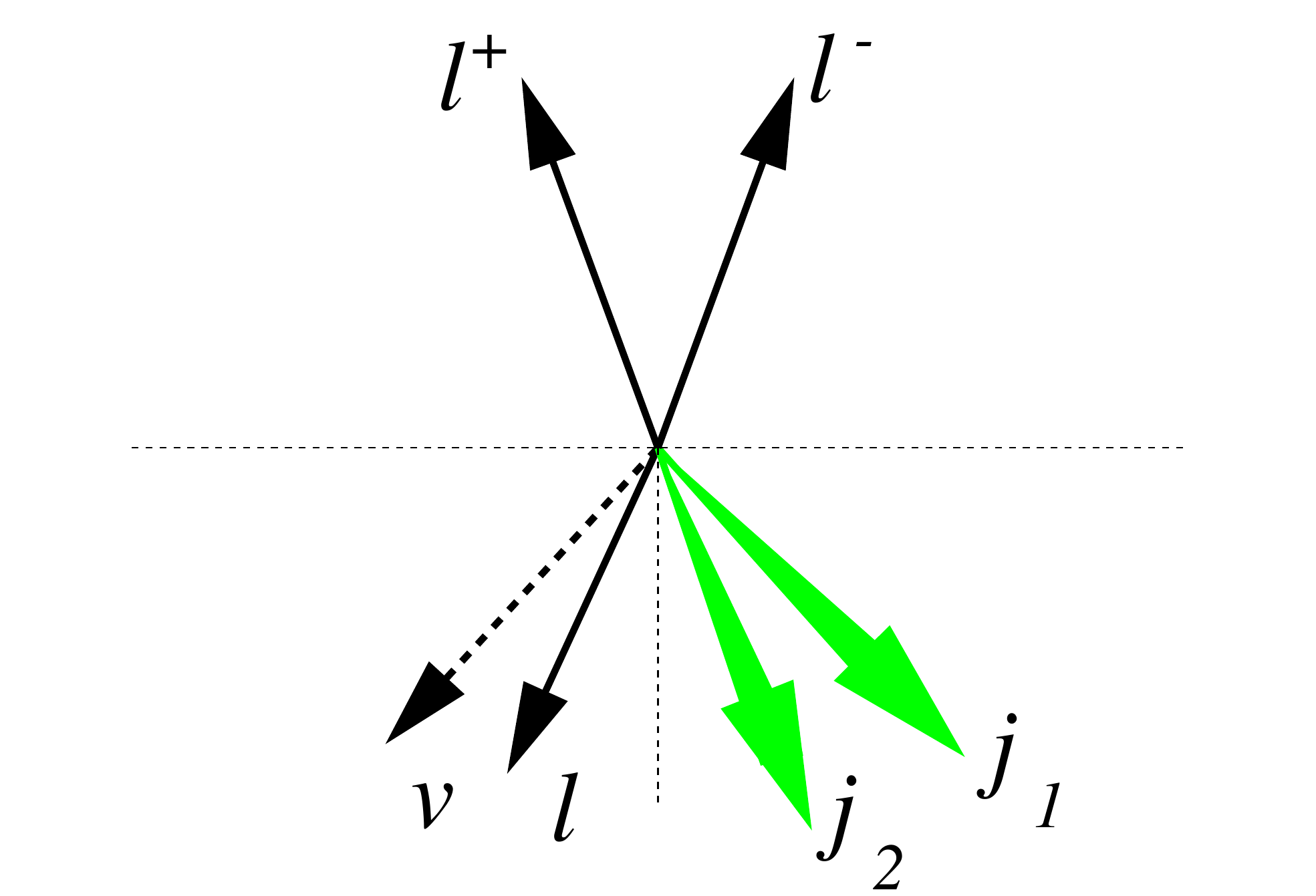}
\put(-21, 8){\textbf{(6)}}\\
\caption{The six signal regions defined in the $3\ell$ channel. The green (red) arrow denotes a normal (fat) jet.}
\label{fig:reg_3lep}
\end{figure}

In the $3\ell$ channel, six signal regions are defined as shown in Tab. \ref{tab:reg_3lep}, and schematically displayed in Fig. \ref{fig:reg_3lep}. Regions (1-3) are characterized by a $W$($\ell\nu$) boson recoiling against a heavy Higgs ($\ell\nu$ denotes the combined 4-vector of $\ell$ and $\nu$ from $W$), from which a boson decays into dilepton, and the other forms a normal jet, a fat jet or two normal jets ($j_{12}$ denotes the combined 4-vector of $j_1$ and $j_2$). Regions (4-6) are similar to (1-3), except that the roles of $W\to\ell\nu$ and $Z\to\ell\ell$ are swapped. The three leptons should have a net charge of $\pm 1$. For $3e$ and $3\mu$ final states, the opposite-charged lepton pair with a smaller $\Delta R$ is regards as from $Z\to\ell\ell$, while for $ee\mu$ and $\mu\mu e$, the correct combination is obvious. To suppress the fake leptons from jets (not modeled in this work) which generally have low $p_T$, the lepton not coming from $Z\to \ell\ell$ is required to have $p_T>50$ GeV. 

The momentum of neutrino from $W\to\ell\nu$ is calculated from the $E_T^\text{miss}$ vector and $W$ mass constraint. When the transverse mass $m_T\ge m_W$\footnote{The transverse mass of a $W$ boson is calculated as $m_T=\sqrt{2p_T^\ell E_T^\text{miss} [1-\cos\Delta\phi(\ell,E_T^\text{miss})]}$.} which happens only because of the $E_T^\text{miss}$ resolution and $W$ boson width effect, $p_z^\nu$ is calculated as $E_T^\text{miss}\cdot p_z^\ell/p_T^\ell$. In the case that $m_T< m_W$, the solutions to $p_z^\nu$ are two-fold. The solution with the a smaller absolute value of $|p_z^\nu|$ is chosen.

For signal with $m_H=300$ GeV or $m_H=900$ GeV, the region definitions are not changed, except that the boson $p_T$ and mass window cuts are adjusted, and the fractions of signal events in different regions will be different.

In all the channels, events are sequentially selected following the order of region numbers (only events not present in the previous region are selected in the next), so there is no overlap between different regions. 

\begin{table}[!hbt]
\centering
\caption{The signal region definitions for $m_H=600$ GeV in the $3\ell$ channel.}
\label{tab:reg_3lep}
\begin{tabular}{l | l}\specialrule{1.0pt}{1pt}{1pt}
 region (1) & region (4) \\ \hline
$p_T^{\ell\nu}>600$ GeV & $p_T^{\ell\ell}>600$ GeV \\ \hline
\multicolumn{2}{c}{$80~\text{GeV}<m_{\ell\ell}<100~\text{GeV}$,}\\
\multicolumn{2}{c}{$60~\text{GeV}< m_{j_1} <160~\text{GeV}$, ~$\tau_2^{j_1}/\tau_1^{j_1} < 0.60$} \\ \hline
$\Delta R(\ell\ell,j_1)<\Delta R(\ell\nu,\ell\ell)$, & $\Delta R(\ell\nu,j_1)<\Delta R(\ell\ell,\ell\nu)$, \\
$\Delta R(\ell\ell,j_1)<\Delta R(\ell\nu,j_1)$, & $\Delta R(\ell\nu,j_1)<\Delta R(\ell\ell,j_1)$, \\
$p_T^{\ell\ell+j_1}>600$ GeV & $p_T^{\ell\nu+j_1}>600$ GeV \\
\specialrule{1.0pt}{1pt}{1pt}
 region (2) & region (5) \\ \hline
$p_T^{\ell\nu}>600$ GeV & $p_T^{\ell\ell}>600$ GeV \\ \hline
\multicolumn{2}{c}{$80~\text{GeV}<m_{\ell\ell}<100~\text{GeV}$,}\\
\multicolumn{2}{c}{$70~\text{GeV}< m_{J} <140~\text{GeV}$, ~$\tau_2^{J}/\tau_1^{J} < 0.50$} \\ \hline
$\Delta R(\ell\ell,J)<\Delta R(\ell\nu,\ell\ell)$, & $\Delta R(\ell\nu,J)<\Delta R(\ell\ell,\ell\nu)$, \\
$\Delta R(\ell\ell,J)<\Delta R(\ell\nu,J)$, & $\Delta R(\ell\nu,J)<\Delta R(\ell\ell,J)$, \\
$p_T^{\ell\ell+J}>600$ GeV & $p_T^{\ell\nu+J}>600$ GeV \\
\specialrule{1.0pt}{1pt}{1pt}
 region (3) & region (6) \\ \hline
$p_T^{\ell\nu}>600$ GeV & $p_T^{\ell\ell}>600$ GeV \\ \hline
\multicolumn{2}{c}{$80~\text{GeV}<m_{\ell\ell}<100~\text{GeV}$,}\\
\multicolumn{2}{c}{$60~\text{GeV}< m_{j_{12}} <120~\text{GeV}$} \\ \hline
$\Delta R(\ell\ell,j_{12})<\Delta R(\ell\nu,\ell\ell)$, & $\Delta R(\ell\nu,j_{12})<\Delta R(\ell\ell,\ell\nu)$, \\
$\Delta R(\ell\ell,j_{12})<\Delta R(\ell\nu,j_{12})$, & $\Delta R(\ell\nu,j_{12})<\Delta R(\ell\ell,j_{12})$, \\
$p_T^{\ell\ell+j_{12}}>600$ GeV & $p_T^{\ell\nu+j_{12}}>600$ GeV \\
\specialrule{1.0pt}{1pt}{1pt}
\end{tabular}
\end{table}

\subsection{Event yields and mass spectrum}

\begin{figure}[!htb]
\centering
\includegraphics[width=0.45\textwidth]{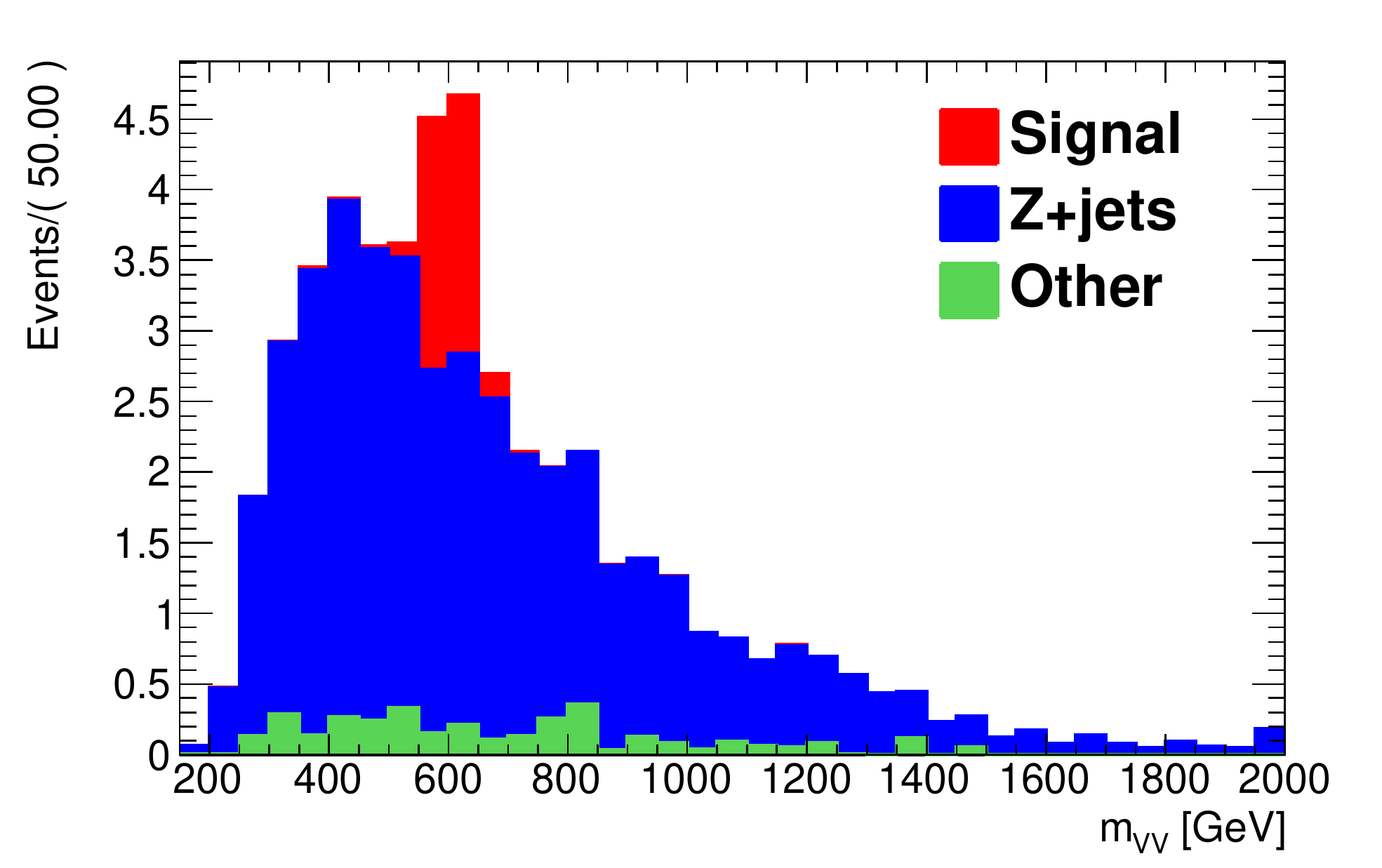}\\
\includegraphics[width=0.45\textwidth]{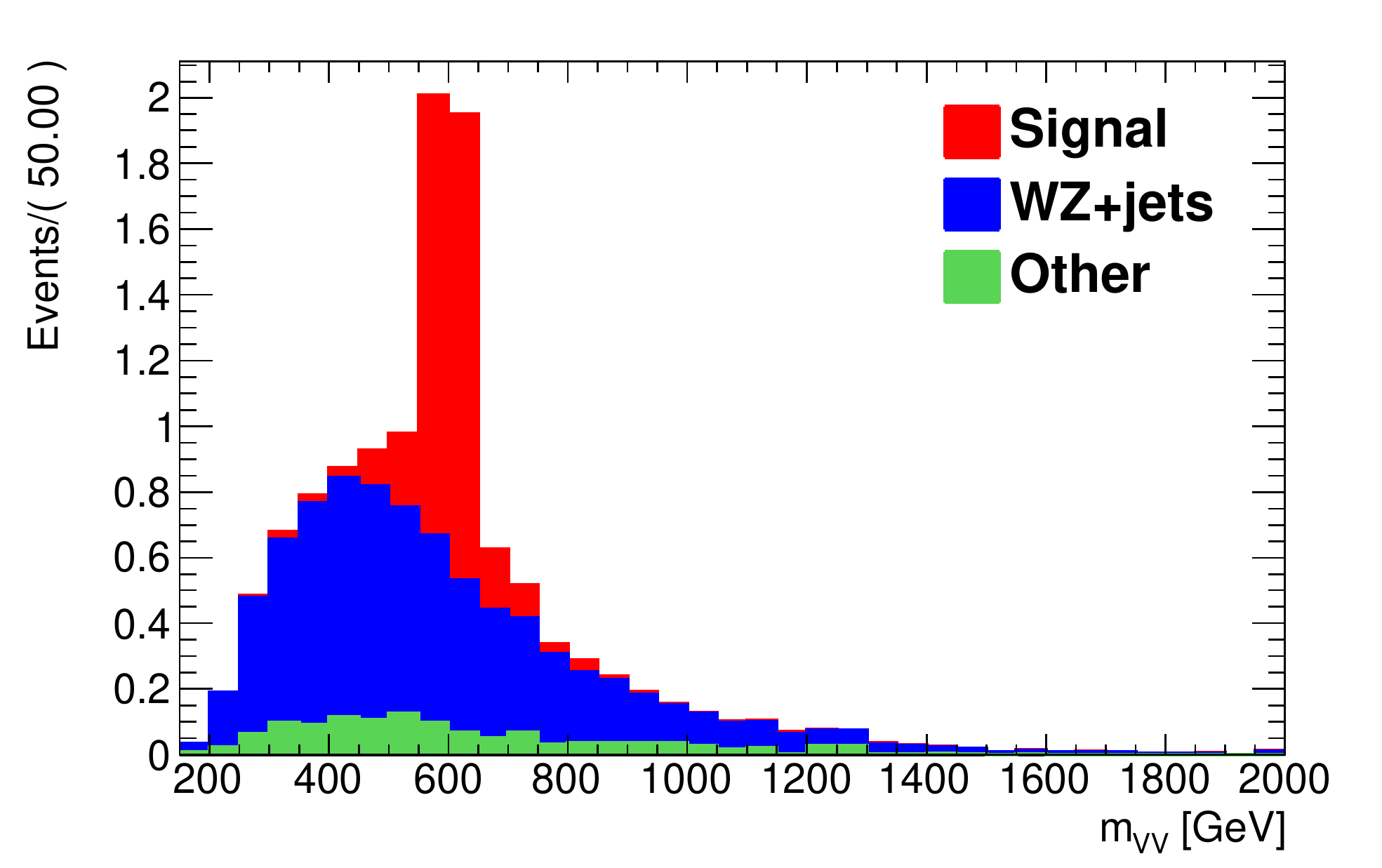}\\
\includegraphics[width=0.45\textwidth]{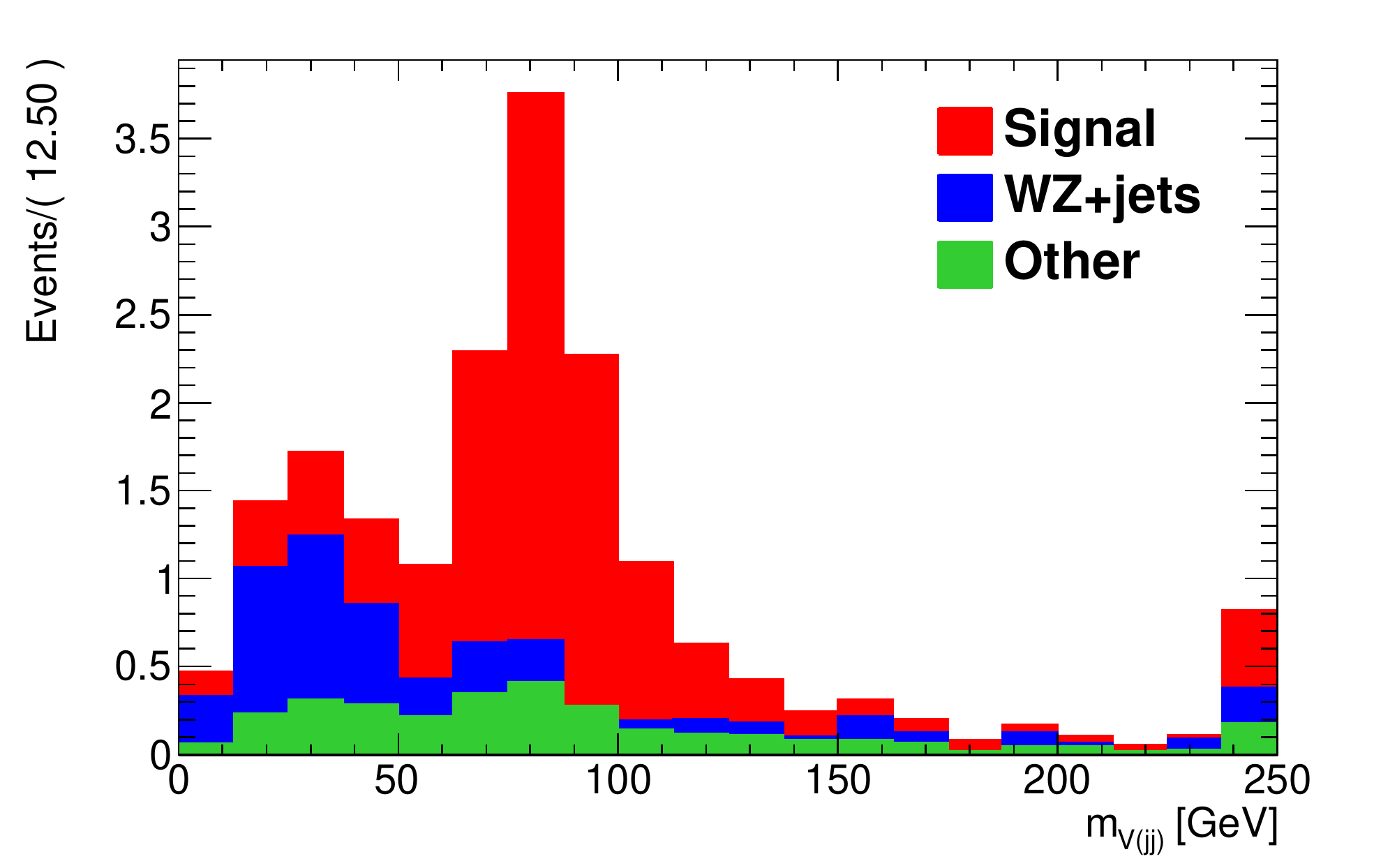}
\caption{The distributions of the heavy Higgs candidate mass in the $2\ell$ OS (top) and $3\ell$ (middle) channel, and the hadronic $V$ boson mass in the $2\ell$ SS  channel (bottom), expected with $300~\text{fb}^{-1}$ of LHC data. In the $2\ell$ channels, ``other" includes $VV$+jets, triboson, $t\bar{t}$ and $t\bar{t}V$ backgrounds, while in the $3\ell$, it includes triboson and $t\bar{t}V$ backgrounds. The signal shown has the following parameters: $m_H=600$ GeV, $\rho_H=0.05$, $f_W=f_{WW}=50$.}
\label{fig:mVV}
\end{figure}

The event yields in the three channels are given in Tab. \ref{tab:yields}. The mass distributions with all signal regions combined in each channel are shown in Fig. \ref{fig:mVV} with the benchmark signal B as in Fig. \ref{fig:vbf_sig}-\ref{fig:vh_sig}. Good heavy Higgs candidate mass can be reconstructed in the 2$\ell$ OS and 3$\ell$ channels, while in the 2$\ell$ SS channel, only the hadronic $V$ boson mass can be shown. The right tail in the signal in the bottom plot of Fig. \ref{fig:mVV} is due to the mis-matched jets illustrated in Fig. \ref{fig:reg_2lep_SS}(3), and a small $W$ boson mass peak is also visible in the ``other" component (dominated by $t\bar{t}V$) of the background. As evident from Tab. \ref{tab:yields}, the 2$\ell$ SS channel provides the best sensitivity among all channels.

\begin{table}[!hbt]
\centering
\caption{The signal and background event yields expected with $300~\text{fb}^{-1}$ of LHC data, with all signal regions combined in each channel. In the $2\ell$ channels, ``other" includes $VV$+jets, triboson, $t\bar{t}$ and $t\bar{t}V$ backgrounds, while in the $3\ell$, it includes triboson and $t\bar{t}V$ backgrounds. The signal shown has the following parameters: $m_H=600$ GeV, $\rho_H=0.05$, $f_W=f_{WW}=50$. }
\label{tab:yields}
\begin{tabular}{llll}\hline
 & Signal & $Z$+QCD jets & Other \\ 
$2\ell$ OS chan. & 4.0 & 41.6 & 3.5 \\ \hline
 & Signal & $WZ$+QCD jets & Other \\ 
$3\ell$ chan. & 3.6 & 7.2 & 1.3 \\ \hline
 & Signal & $WZ$+QCD jets & Other \\ 
$2\ell$ SS chan. & 11.4 & 4.1 & 3.1 \\
\hline
\end{tabular}
\end{table}

\section{Sensitivity in the model parameter space}

To extract the signal sensitivity, mass window cuts are applied to the distributions shown in Fig. \ref{fig:mVV}, and number counting is performed. The sensitivity is based on ratios of Poisson likelihoods, and toy distributions are obtained for background-only and signal+background hypotheses.
In this work, three mass parameters: $m_H=300,~600,~900$ GeV, are investigated. 
%
%
Since $\rho_h\simeq 1$ from current Higgs measurement, it is expected that $\rho_H$ is not very large. Hence $\rho_H=0.05$ is taken as a benchmark value, and the two dimensional parameter space of $f_W$ and $f_{WW}$ is scanned. Since the Higgs width is proportional to $\rho_H^2 m_H^3$, the small $\rho_H$ also makes the Higgs width small. With $\rho_H=0.05$ and $f_W=f_{WW}=50$, a Higgs of mass 900 GeV has a width of only 0.571 GeV. Therefore, the interference between the signal and the SM triboson background (since they have the same final state) can be safely neglected. 

\begin{figure}[!tb]
\centering
\includegraphics[width=0.40\textwidth]{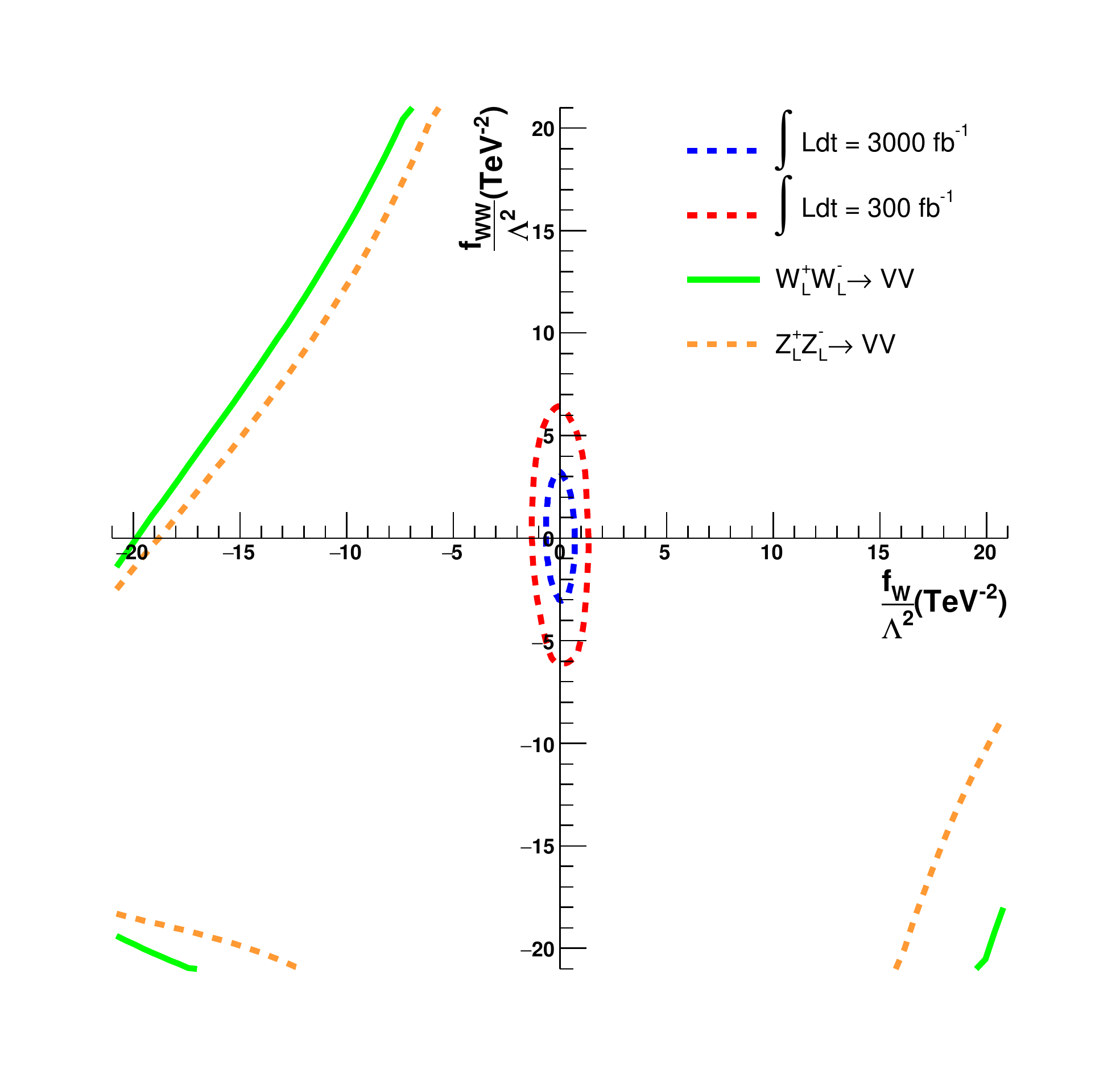}
\caption{The excluded regions at $95\%$ CL in the $f_W - f_{WW}$ parameter space for $\mathcal{L}=300~\text{fb}^{-1}$ and $3~\text{ab}^{-1}$ scenarios, with $\rho_H=0.05$ and $m_H=300$ GeV. The unitarity bounds from gauge boson scattering are also shown. }
\label{fig:exclusion_300}
\end{figure}

\begin{figure}[!tb]
\centering
\includegraphics[width=0.40\textwidth]{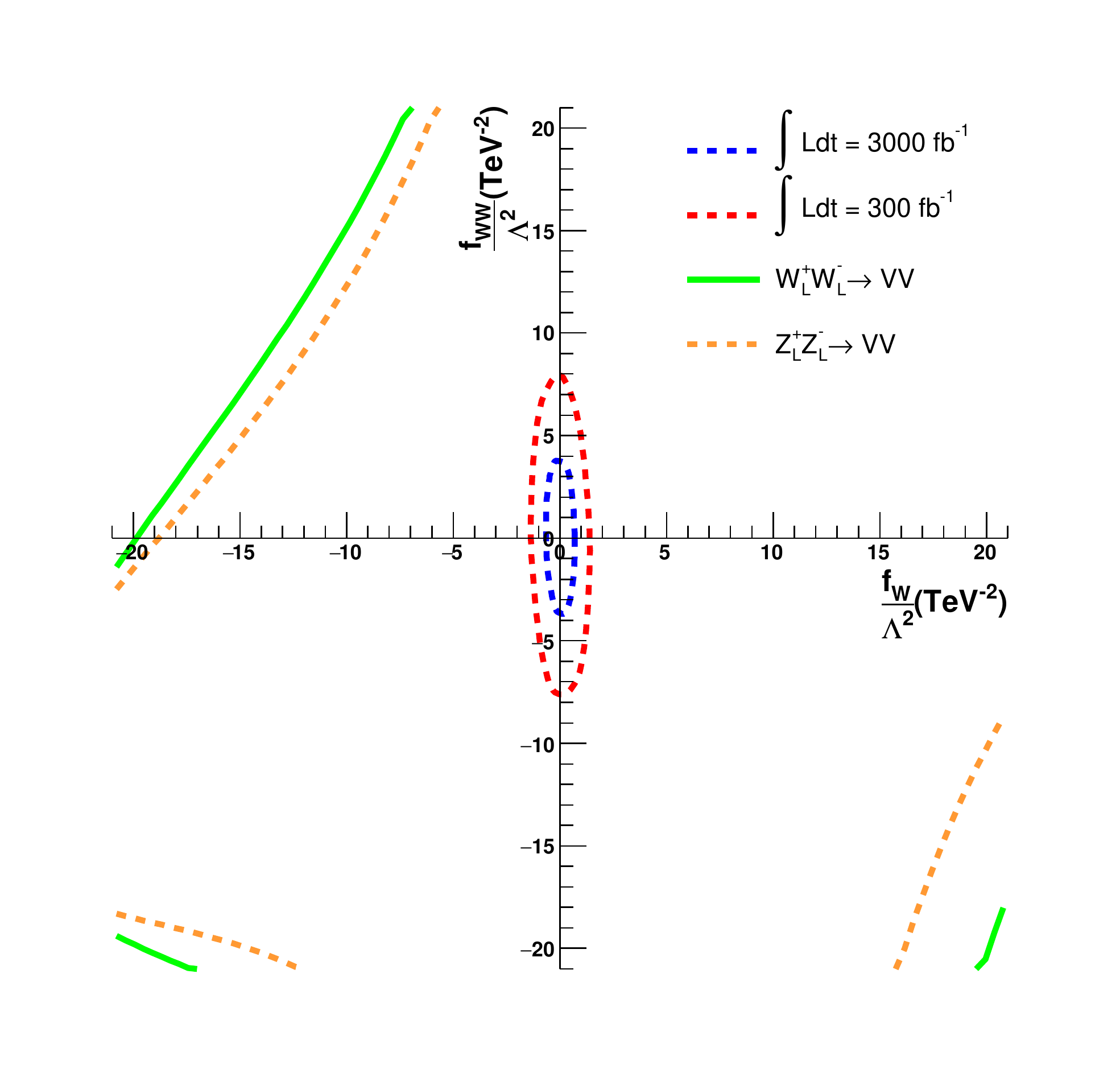}
\caption{The excluded regions at $95\%$ CL in the $f_W - f_{WW}$ parameter space for $\mathcal{L}=300~\text{fb}^{-1}$ and $3~\text{ab}^{-1}$ scenarios, with $\rho_H=0.05$ and $m_H=600$ GeV. The unitarity bounds from gauge boson scattering are also shown. }
\label{fig:exclusion_600}
\end{figure}

\begin{figure}[t]
\centering
\includegraphics[width=0.40\textwidth]{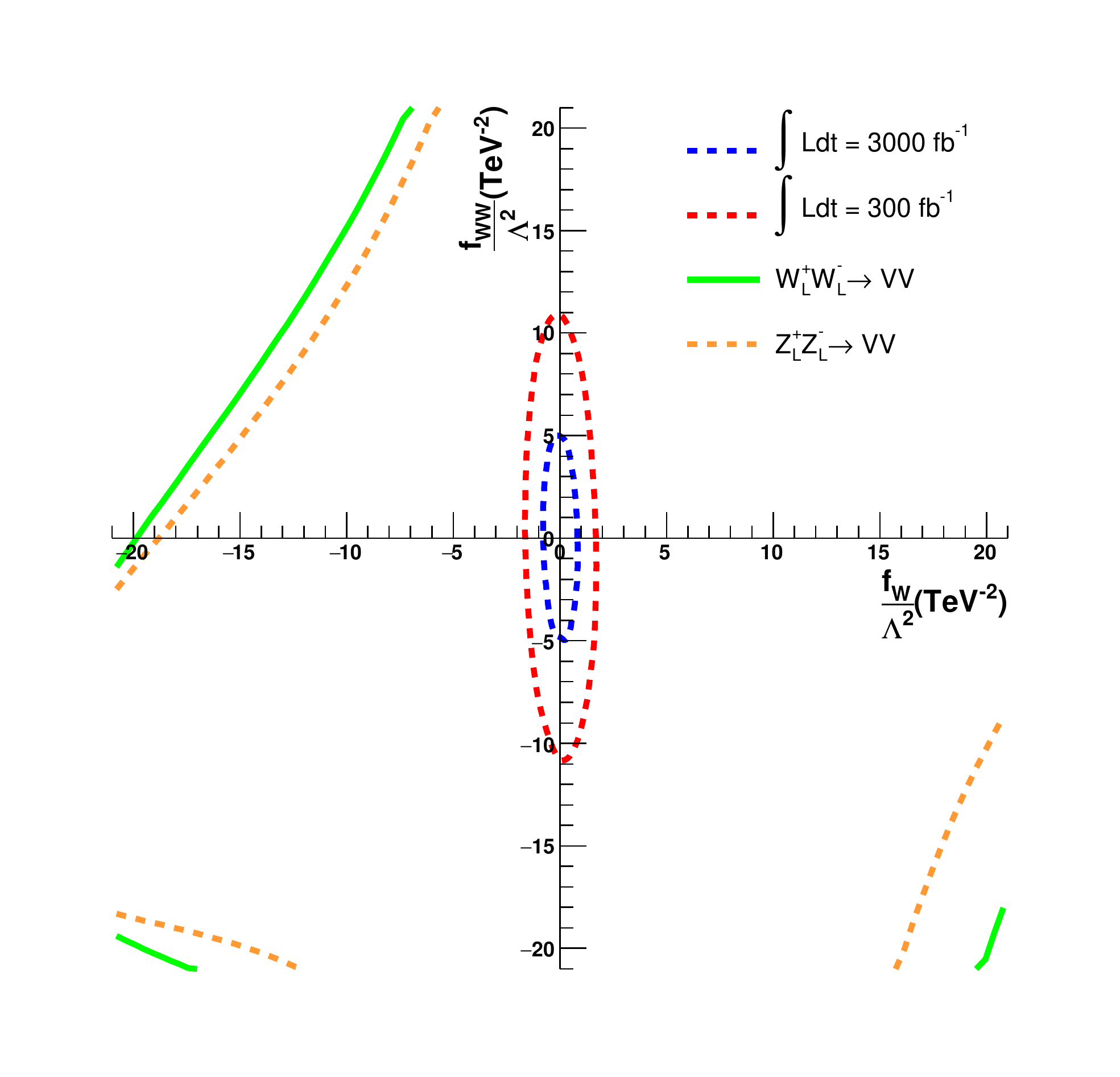}
\caption{The excluded regions at $95\%$ CL in the $f_W - f_{WW}$ parameter space for $\mathcal{L}=300~\text{fb}^{-1}$ and $3~\text{ab}^{-1}$ scenarios, with $\rho_H=0.05$ and $m_H=900$ GeV. The unitarity bounds from gauge boson scattering are also shown. }
\label{fig:exclusion_900}
\end{figure}

Suppose there is no heavy Higgs signal with large dim-6 operator coefficients, the $95\%$ Confidence Level (CL) exclusion regions for three different Higgs masses are shown in Fig. \ref{fig:exclusion_300}-\ref{fig:exclusion_900}, for two integrated data luminosities: $300~\text{fb}^{-1}$ and $3~\text{ab}^{-1}$, combining the $2\ell$ and $3\ell$ channels. The bounds based on the consideration of gauge boson scattering amplitude unitarity \cite{Kuang} are also shown in these figures. It is evident that a large part of the parameter space allowed by unitarity can be excluded, with just $300~\text{fb}^{-1}$ of data. It is worthwhile to note that with $\rho_h=1$ large values of $\rho_H$ will shift the area enclosed by the unitarity bounds away from the origin, making these signals much easier to be excluded.

\section{Conclusion}

In summary, a search strategy for a heavy Higgs with generic dim-6 couplings to SM gauge boson is presented in this work. We go beyond the final state studied in Ref. \cite{Kuang} to focus on the two and three lepton final states, where the SM background can be substantially suppressed by means of boosted boson jets and jet substructure moments. The signal we are looking at can be sparse in ggF and VBF productions (thus escaped detection so far), but can be found in the VH production with proper sets of cuts. This is a phase space corner not touched upon by LHC up to date, and searching for such a generic heavy Higgs may shed light on something new toward BSM.

\section*{Acknowledgments}
X. Chen is supported by the National Key Research and Development of China (Grant. No. 2018YFA0404002) and Tsinghua University Initiative Scientific Research Program. Y. Wu is supported by the Natural Sciences and Engineering Research Council of Canada (NSERC). Y.-P. Kuang is supported by the National Natural Science Foudation of China (grant 11135003 and 11275102). Q. Wang is supported by the National Key Research and Development of China (Grant. No. 2017YFA0402201). S.-C. Hsu is supported by the DOE Office of Science, Office of High Energy Physics Early Career Research program under Award Number DE-SC0015971. Z. Hu is supported by the Tsinghua University Initiative Scientific Research Program.


\bibliographystyle{elsarticle-num}
\bibliography{references}



\end{document}